%% file: main.tex
\documentclass[]{spie}  

\usepackage{amsmath,amsfonts,amssymb}
\usepackage{graphicx}
\usepackage[colorlinks=true, allcolors=blue]{hyperref}

\usepackage{float}
\usepackage{caption}
\usepackage{subcaption}
\title{Off-axis Hartmann Wavefront Sensing for the GMT-Consortium Large Earth Finder (G-CLEF) Red Camera Optics}

\author[a,b]{Matthew C. H. Leung}
\author[a,c]{Colby A. Jurgenson}
\author[a,b,c]{Andrew Szentgyorgyi}
\author[a,c]{Brian McLeod}
\author[a,c]{Cem Onyuksel}
\author[a,c]{Joseph Zajac}
\author[a,b]{David Charbonneau}
\author[a,c]{William Podgorski}
\author[a,c]{Abigail Unger}
\author[a,c]{Mark Mueller}
\author[a,c]{Matthew Smith}
\author[a,c]{Daniel Baldwin}
\author[a,b]{V. Ashley Villar}

\affil[a]{Center for Astrophysics \textbar{} Harvard \& Smithsonian, 60 Garden St, Cambridge, MA 02138, USA}
\affil[b]{Department of Astronomy, Harvard University, 60 Garden St, Cambridge, MA 02138, USA}
\affil[c]{Smithsonian Astrophysical Observatory, 100 Acorn Park Dr, Cambridge, MA 02140, USA}

\authorinfo{Further author information: (Send correspondence to M.C.H. Leung)\\M.C.H. Leung: E-mail: matthew.leung@cfa.harvard.edu \\  C.A. Jurgenson: E-mail: colby.jurgenson@cfa.harvard.edu}

\begin{document} 
\maketitle

\begin{abstract}
The Hartmann test is a method used to measure the wavefront error in a focal optical system, wherein a mask with a pattern of small holes is placed at the system’s aperture stop. By taking an image at a defocused plane, the differences between the ideal and real positions of the reimaged holes (called the transverse ray aberrations) can be measured, which can then be used to estimate the wavefront error. However, the Hartmann test is usually used with an on-axis field. In this paper, we present a wavefront sensing method which generalizes the classical Hartmann test for off-axis field angles and arbitrary reference wavefronts. Our method involves taking images at two defocused planes, and then using the real reimaged hole positions on both planes to estimate the trajectories of rays from the system’s exit pupil, at which the reference wavefront is situated. We then propagate the rays forward from the reference wavefront to one of the two defocused planes, in order to find the ideal reimaged hole positions, from which we can compute transverse ray aberrations. We derive and solve a pair of nonlinear partial differential equations relating transverse ray aberrations to wavefront error, using Zernike decomposition and nonlinear least squares. Our method has been verified on simulated data from the 7-lens $f$/2.25 red camera system of the GMT-Consortium Large Earth Finder (G-CLEF), a high resolution optical echelle spectrograph which will be a first light instrument for the Giant Magellan Telescope (GMT).
\end{abstract}



\input{Sections/intro}
\input{Sections/red_cam}
\input{Sections/classical_hartmann}
\input{Sections/app}
\input{Sections/method_overview}
\input{Sections/TRA}
\input{Sections/TRA_to_WFE}
\input{Sections/results}

\input{Sections/conclusions}
\input{Sections/appendix}

\acknowledgments 
 
M.C.H. Leung gratefully acknowledges support from the Dr. Gerald A. Soffen Memorial Fund and NASA Academy Alumni Association.

\bibliography{report} 
\bibliographystyle{spiebib} 

\end{document}

%% file: Sections/intro.tex
\section{Introduction}\label{sec:intro}

The classical Hartmann test is a wavefront sensing method for focal optical systems\cite{Hartmann1900,MalacaraOST}. In this method, a mask with a pattern of small holes is placed at the optical system's aperture stop. This mask allows for transverse ray aberrations to be measured at a defocused plane, from which the wavefront error can be estimated. However, the classical Hartmann test is usually used with an on-axis field, and practical implementations of this method can fail for off-axis field angles and fast beams. In this paper, we present a wavefront sensing method which generalizes the classical Hartmann test for off-axis field angles. Our method also works with fast beams, arbitrary reference wavefronts, and irregular obscurations. We developed this method to evaluate the off-axis performance of the 7-lens $f$/2.25 red camera system of the GMT-Consortium Large Earth Finder (G-CLEF\cite{Szentgyorgyi2018,Szentgyorgyi2016,Szentgyorgyi2024}), a high resolution optical echelle spectrograph for the Giant Magellan Telescope (GMT) and the Magellan Clay Telescope.

An outline of this paper is as follows. In Section \ref{sec:red_cam}, we discuss the design of G-CLEF and an experimental test setup used to evaluate the performance of the G-CLEF red camera. In Section \ref{sec:background}, we discuss the challenges and restrictions associated with using certain wavefront sensing methods to measure the off-axis wavefront error in the G-CLEF red camera using our test setup. We discuss the shortcomings of the classical Hartmann test and motivate the need to develop a generalized method for off-axis field angles. In Section \ref{sec:Hartmann_app}, we discuss the Hartmann test apparatus we used with our test setup. In Sections \ref{sec:overview}, \ref{sec:TRA}, and \ref{sec:TRA_to_WFE}, we discuss our generalized method, which has two parts: measuring transverse ray aberrations using images taken on two defocused planes (\mbox{Section \ref{sec:TRA}}), and estimating the wavefront error using the transverse ray aberration measurements (Section \ref{sec:TRA_to_WFE}). In Section \ref{sec:results}, we present the results of our method applied to simulated data from the G-CLEF red camera.

%% file: Sections/red_cam.tex
\section{G-CLEF and Red Camera Test Setup}\label{sec:red_cam}

\subsection{GMT-Consortium Large Earth Finder (G-CLEF)}

\begin{figure}[htb!]
  \centering
  \includegraphics[width=0.75\textwidth]{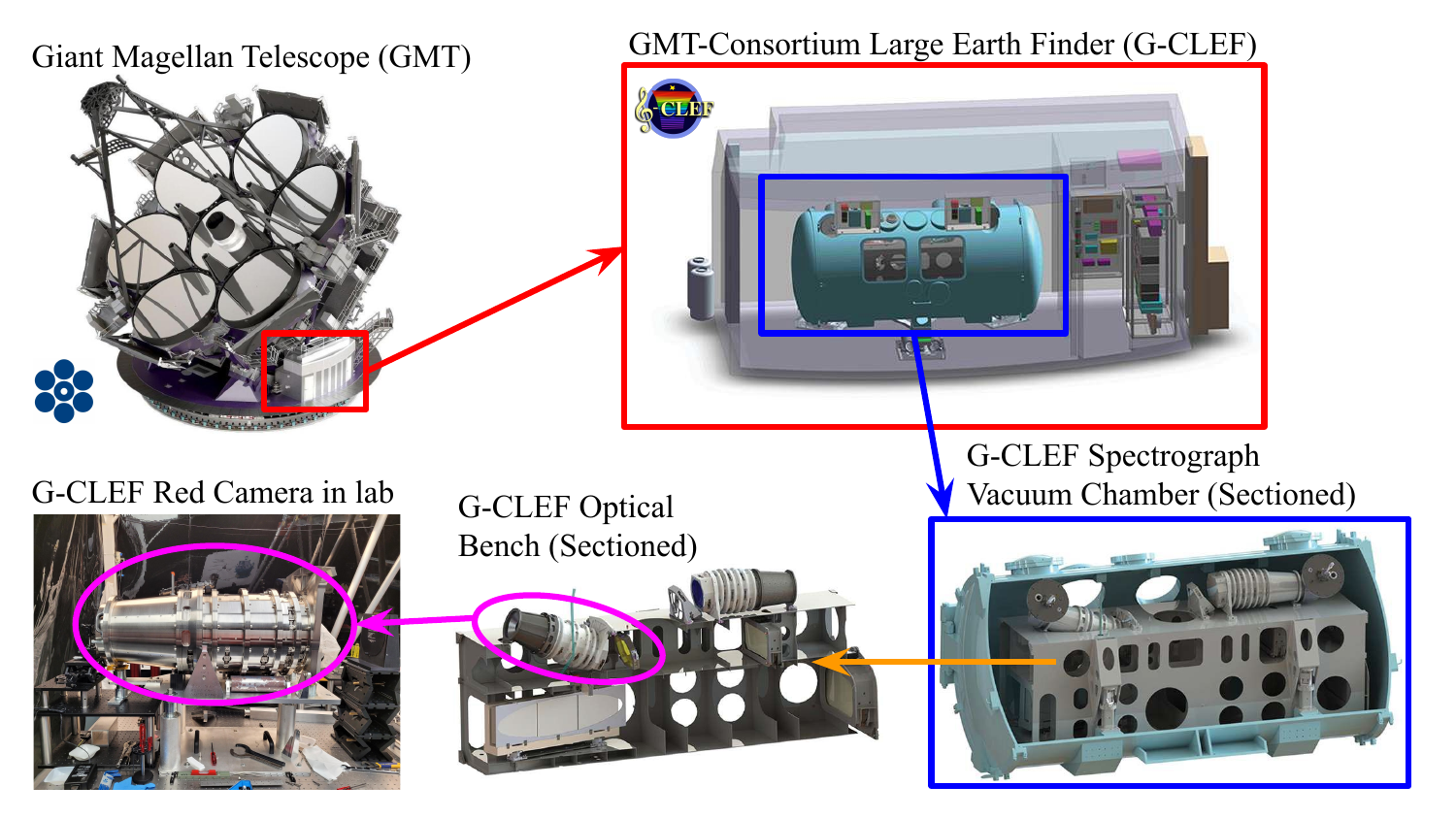}
  \caption{The opto-mechanical design\cite{Mueller2022} of G-CLEF, and the red camera within the instrument.}
  \label{fig:G-CLEF_and_Red}
\end{figure}

\begin{figure}[htb!]
  \centering
  \includegraphics[width=0.65\textwidth]{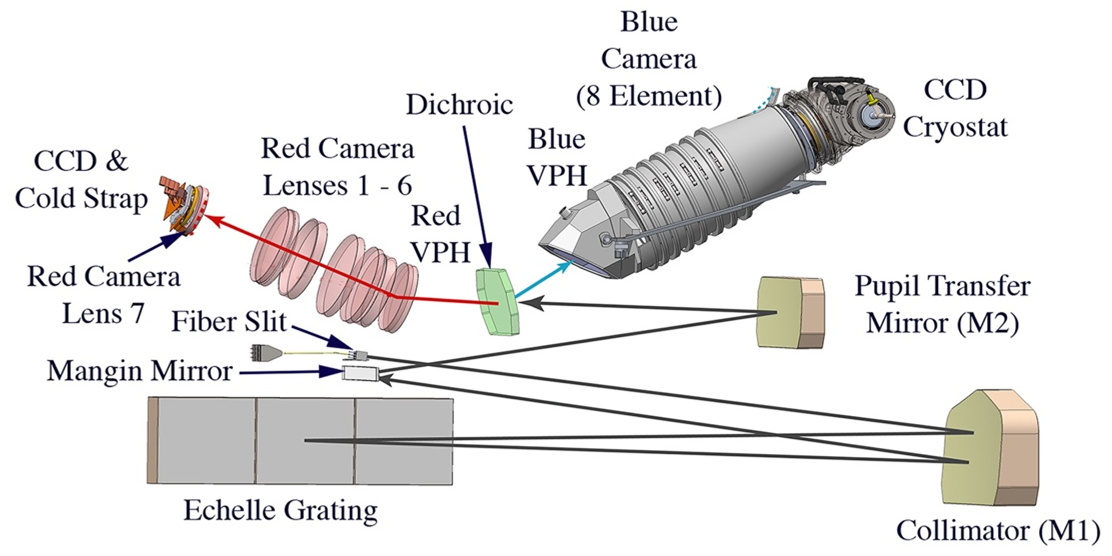}
  \caption{The G-CLEF spectrograph optical component layout\cite{Szentgyorgyi2018,Szentgyorgyi2016,Szentgyorgyi2024}. The red camera is in the upper left part of this illustration. The red camera bezels are not shown here, and only the red camera lenses are shown.}
  \label{fig:G-CLEF_optical}
\end{figure}

G-CLEF is a precision radial velocity spectrograph which will discover and characterize Earth-like planets in the habitable zone of Sun-like stars. Figures \ref{fig:G-CLEF_and_Red} and \ref{fig:G-CLEF_optical} show the opto-mechanical design and optical layout of G-CLEF. Light is injected into G-CLEF via an optical fiber and is collimated by a mirror M1. The collimated light is dispersed by a reflective echelle grating, and eventually passes through a dichroic, which splits the light into a blue channel (350 nm to 540 nm) and a red channel (540 nm to 950 nm). Each channel includes a volume phase holographic (VPH) grism as a cross-disperser, and also a camera to image the spectrum onto a CCD detector.

At the time of writing, G-CLEF is currently being assembled at the Center for Astrophysics \textbar{} Harvard \& Smithsonian. In 2027, G-CLEF is planned to be commissioned on the 6.5m Magellan Clay Telescope at the Las Campanas Observatory in Chile. In the 2030s, G-CLEF is planned to be commissioned as a first light instrument on the GMT. More information about G-CLEF can be found in other papers published in these proceedings\cite{Szentgyorgyi2024, Jurgenson2024, Mueller2024, Lupinari2024, Sofer-Rimalt2024, May2024, Millan-Gabet2024, Zafar2024}.


\subsection{G-CLEF Red Camera Test Setup}\label{sec:red_cam_test}

\begin{figure}[H]
  \centering
  \includegraphics[width=0.9\textwidth]{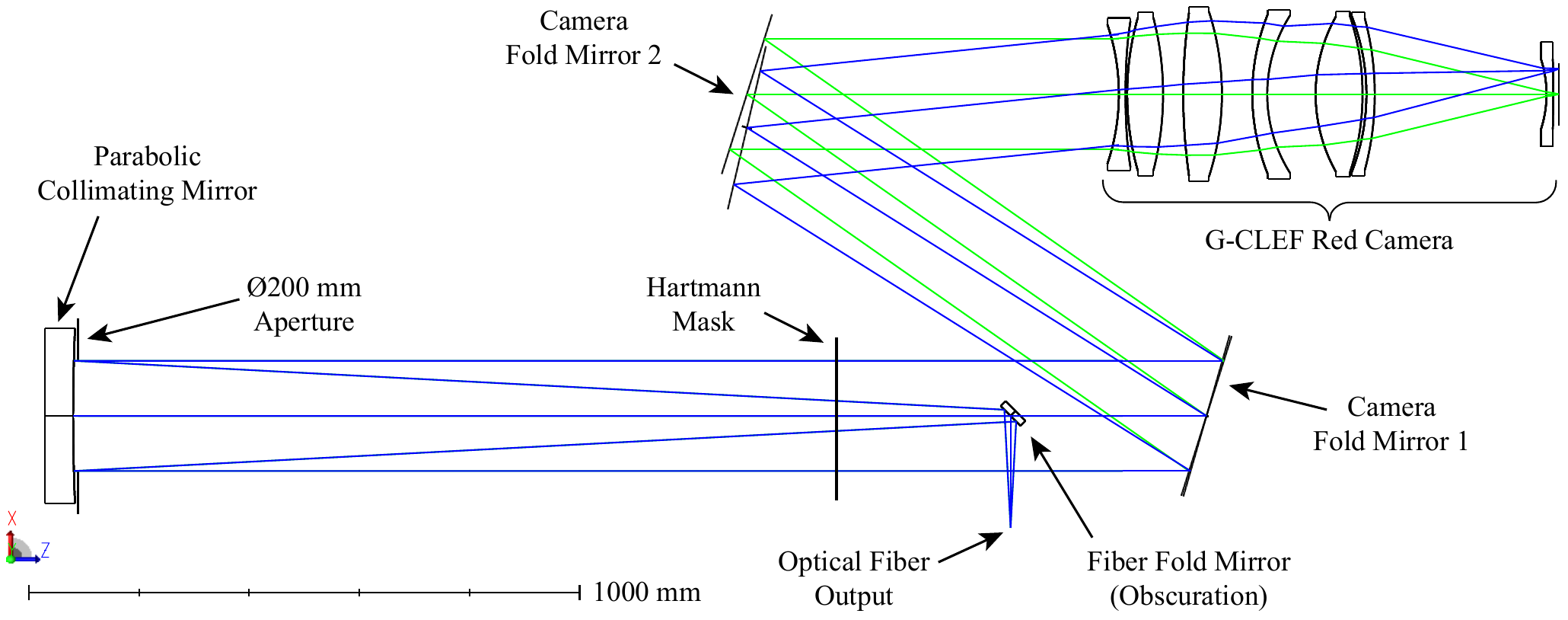}
  \caption{G-CLEF red camera test setup. A parabolic mirror sends collimated monochromatic light towards two camera fold mirrors, which inject light into the red camera at the same trajectories as if G-CLEF's echelle grating and cross-disperser were present. In this figure, the green rays correspond to on-axis light, and the blue rays correspond to G-CLEF's $m=114$ diffraction order at $\lambda=536$ nm.}
  \label{fig:Red_AIT_m114}
\end{figure}

\begin{figure}[H]
  \centering
  \includegraphics[width=0.5\textwidth]{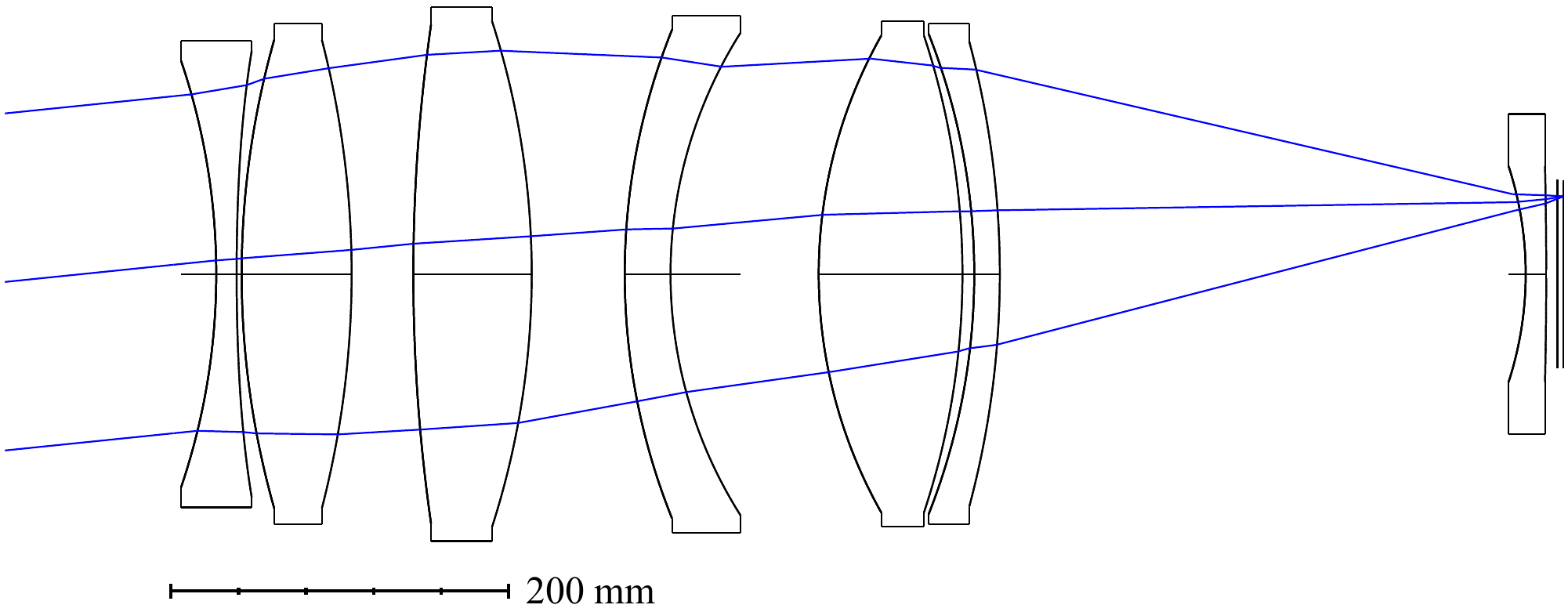}
  \caption{Closeup of the 7-lens $f$/2.25 G-CLEF red camera, with light injected at an off-axis field angle. The trajectory of the rays here correspond to G-CLEF's $m=114$ diffraction order at $\lambda=536$ nm.}
  \label{fig:Red_AIT_standalone}
\end{figure}

\begin{figure}[H]
  \centering
  \includegraphics[width=0.8\textwidth]{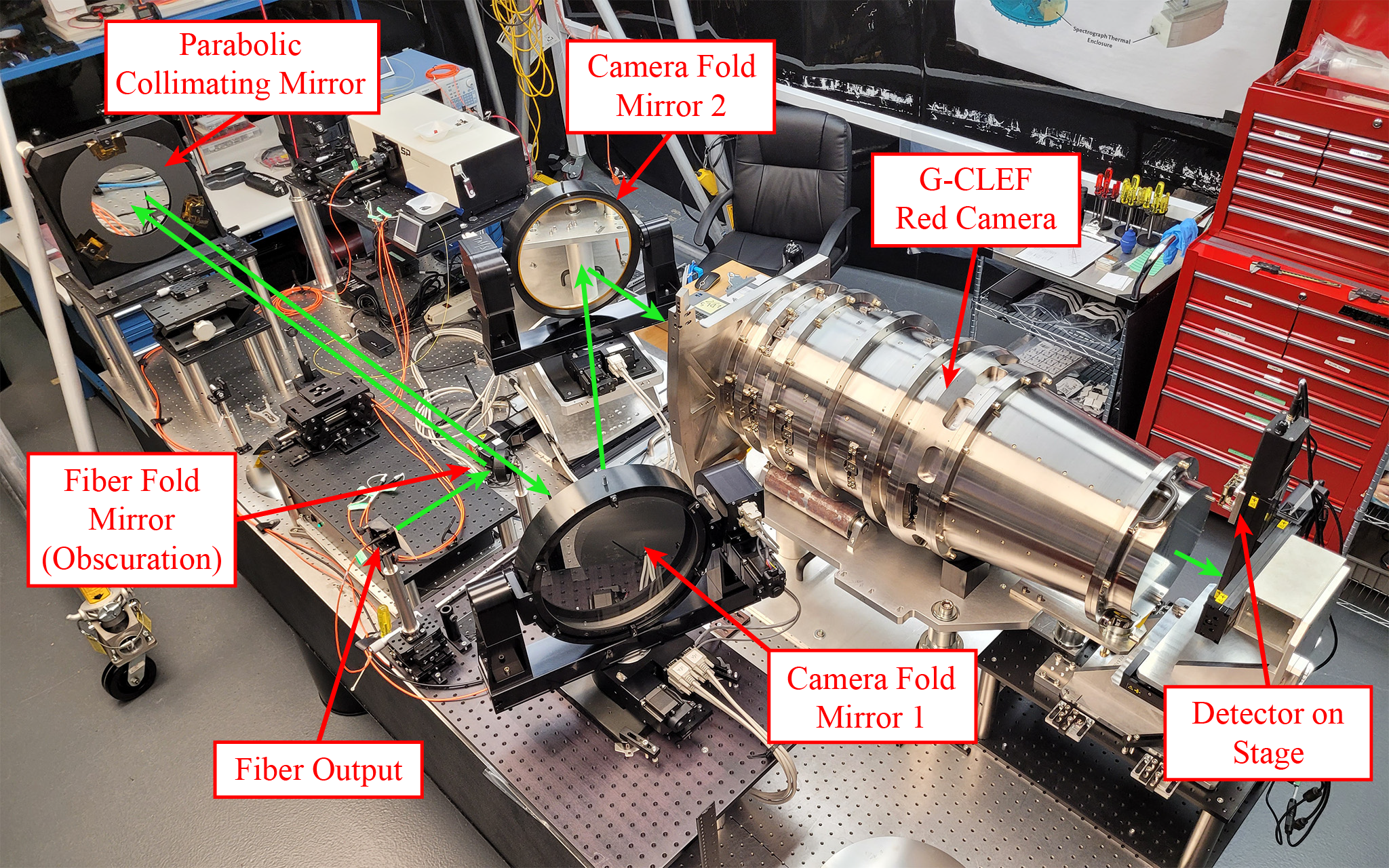}
  \caption{Real-world photo of the test setup in Figure \ref{fig:Red_AIT_m114}.}
  \label{fig:Red_AIT_lab_layout}
\end{figure}

At the time of writing, the G-CLEF red camera has been assembled, and tests are currently being conducted to verify its performance\cite{Jurgenson2024}. In order to test the red camera, we developed a setup shown in Figure \ref{fig:Red_AIT_m114}. An optical fiber injects light into this setup, from a monochromator. A fold mirror then reflects the diverging beam from the fiber to the left, towards a parabolic mirror that reflects the beam towards the right, collimating it. A \mbox{200 mm} diameter aperture stop is at the parabolic mirror. The collimated beam then makes a second pass through the fiber fold mirror, which obscures a part of the beam. The fiber fold mirror is an irregular obscuration (see \mbox{Figure \ref{fig:obs}}), and is a cause of problems which will be discussed in Section \ref{sec:WFS_focal}. The beam is then reflected by two fold mirrors on gimbal mounts and translating stages, and is then injected into the G-CLEF red camera. The two fold mirrors inject monochromatic light into the red camera at the same trajectories as if the echelle grating and VPH cross-disperser were present. In other words, this setup mimics the optical train before the red camera shown in \mbox{Figure \ref{fig:G-CLEF_optical}}. Figure \ref{fig:Red_AIT_standalone} shows a close up of the red camera, with the trajectory of the injected light corresponding to the echelle grating's $m=114$ diffraction order at $\lambda=536$ nm. In addition, we placed a detector on a translating stage near the focus of the camera to take images. This can be seen in Figure \ref{fig:Red_AIT_lab_layout}, which shows the setup in \mbox{Figure \ref{fig:Red_AIT_m114}} implemented in the lab. More details about our red camera test setup, including other tests conducted with our setup, can be found in another paper published in these proceedings\cite{Jurgenson2024}.

Using this setup, we would like to measure the wavefront error of the red camera when light is injected into the red camera at a trajectory corresponding to a particular echelle grating order and wavelength. We would like to quantify particular aberrations (defocus, astigmatism, coma, trefoil, spherical aberration, etc.) in the red camera when off-axis light is injected. However, using this setup gives rise to several challenges which limit the type of wavefront sensing methods we can use, which we will discuss in Section \ref{sec:background}.

%% file: Sections/classical_hartmann.tex
\section{Background: Wavefront Sensing and the Hartmann Test}\label{sec:background}

\subsection{Wavefront Sensing for Focal Optical Systems}\label{sec:WFS_focal}

The wavefront error (also called the ``wavefront aberration function'') is an important metric for the quality of an optical system. The wavefront error is the difference between the system's actual wavefront $\Phi$ and an ideal reference wavefront $\Phi_r$. A wavefront of an electromagentic wave is a surface containing points with equal phase. We define the wavefront error $W$ to be:
\begin{equation}\label{eq:WFE}
    W(x,y) \equiv \Phi(x,y) - \Phi_r(x,y)
\end{equation}
where $x$ and $y$ are physical coordinates on the system's exit pupil plane. Note that wavefront error has physical units of length, but is customarily reported in units of waves (through division by wavelength $\lambda$) or radians (through multiplication by wavenumber $k = 2\pi/\lambda$). In addition, wavefront error is commonly expressed as a linear combination of Zernike circle polynomials for optical systems with a circular aperture (see Appendix \ref{sec:app_Zernike}).

There are many wavefront sensing methods which can be used to measure the wavefront error of an optical system. However, if we want to measure the wavefront error in the G-CLEF red camera using our test setup in Section \ref{sec:red_cam_test}, there are three main challenges with our setup which limit the type of wavefront sensing methods which we can use. These three challenges are: (1) the need for a wavefront sensing method that works with focal systems, (2) the presence of an irregular obscuration in our setup, and (3) the need for a wavefront sensing method that works with off-axis field angles.

The first challenge is that the placement of the test detector stage in our setup prevents us from adding additional apparatus beyond the focus of the red camera. We cannot, for example, place another lens after the red camera's focus to collimate the diverging beam after focus. Hence, we need a wavefront sensing method that works for focal systems, as opposed to a wavefront sensing method that works with afocal systems (where the beam is collimated). For example, we cannot use a Shack-Hartmann wavefront sensor (which is a variant of the Hartmann test\cite{MalacaraHernndez2015}), because we cannot fit in the necessary apparatus.

One method that works with focal optical systems is Curvature Wavefront Sensing (CWFS\cite{Roddier1993}), a method which we initially tried with our test setup. CWFS is one of a series of wavefront sensing methods utilizing the Transport of Intensity Equation (TIE), which is an elliptic partial differential equation (PDE) relating the intensity $I(x,y,z)$ of a monochromatic, coherent, electromagnetic beam (propagating in the $+z$ direction) to its phase $\phi(x,y)$ \cite{Zuo2020}: $-k \frac{\partial}{\partial z} I(x,y) = \nabla I(x,y,z) \cdot \nabla \phi(x,y) + I(x,y,z) \nabla^2 \phi(x,y)$. On the right hand side of the TIE, the first term is called the phase gradient and the second term is called the phase curvature. An assumption is made that the intensity at the pupil plane $I(x,y,0)$ is sufficiently constant on its domain so that the phase gradient is zero \cite{Roddier1993,Xin2015}, and so the phase curvature is the only nonzero term on the right hand side. In CWFS, we take two defocused images, one intrafocal (before the focus) and one extrafocal (after the focus), which are used to approximate $\nabla^2 \phi (x,y)$. The TIE is then reduced to a Poisson equation\cite{Waller2010}, and the wavefront error can be estimated using an iterative application of an inverse Fourier transform\cite{Xin2015}. However, a key assumption of CWFS is that the intensity at the pupil plane is sufficiently constant on its domain. If there is an obscuration, CWFS can still work if we select an appropriate domain for the problem. For example, if there is a circular obscuration, we can select an annulus as the domain, use annular Zernike polynomials to decompose the wavefront error, and iteratively apply an inverse Fourier transform \cite{Xin2015}. However, in the red camera test setup, the fiber fold mirror in Figure \ref{fig:Red_AIT_m114} causes a highly irregular obscuration (see Figure \ref{fig:obs}), such that there is no trivial way to recover the wavefront error using CWFS. We initially tried using CWFS with our test setup, but we were unable to get any good results. The irregular obscuration in our test setup is the second challenge.

One method that works with both focal optical systems and irregular obscurations is the Hartmann test. In Section \ref{sec:classical_Hartmann}, we discuss the classical Hartmann test, and why our third challenge of off-axis field angles motivates the need to develop a more generalized method.


\subsection{The Classical Hartmann Test and its Shortcomings}\label{sec:classical_Hartmann}

\begin{figure}[H]
    \centering
    \begin{subfigure}{0.49\textwidth}
        \centering
        \includegraphics[width=\textwidth]{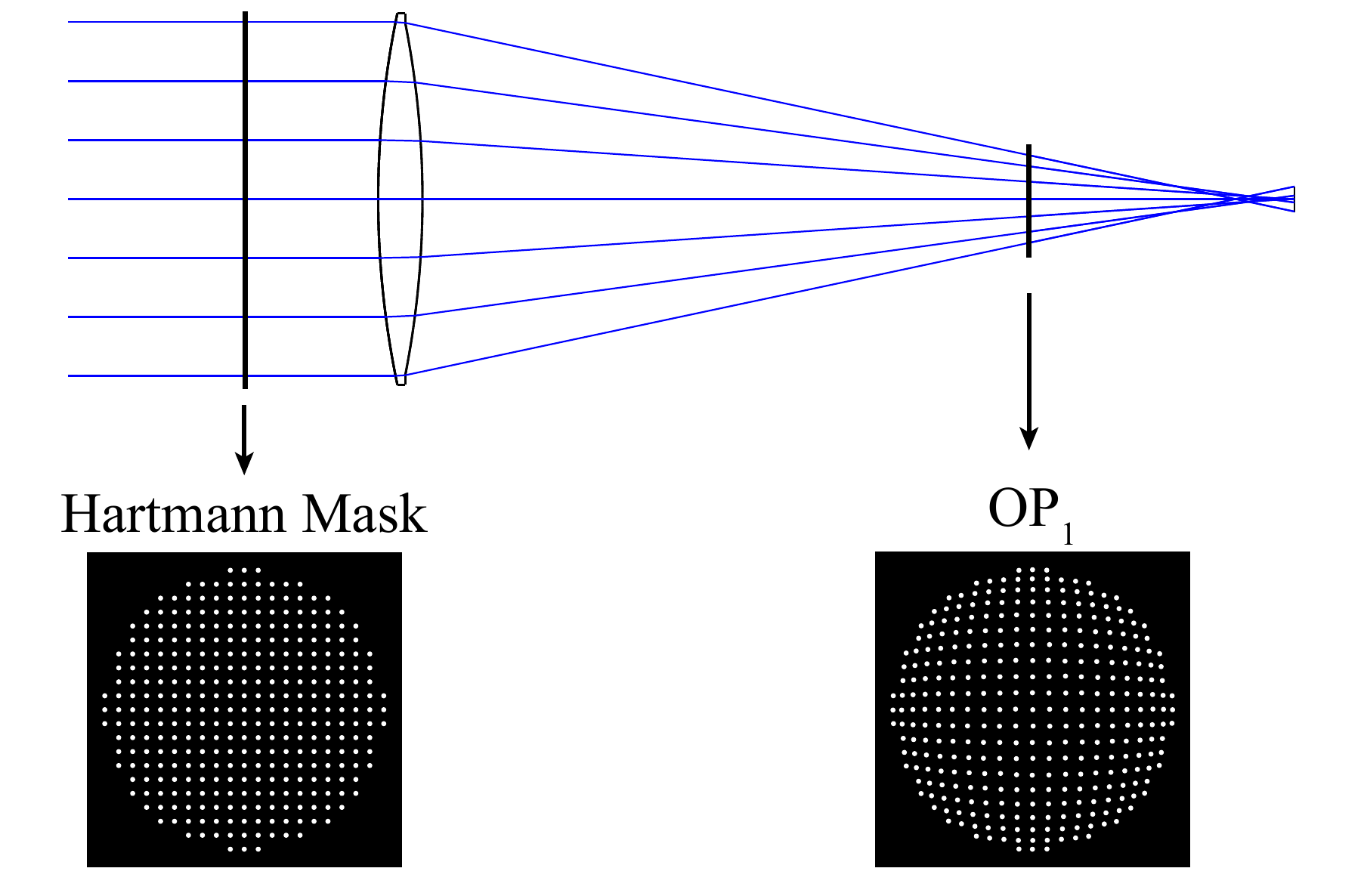}
        \caption{Classical Hartmann test}
        \label{fig:Hartmann_classic}
    \end{subfigure}
    \begin{subfigure}{0.49\textwidth}
        \centering
        \includegraphics[width=\textwidth]{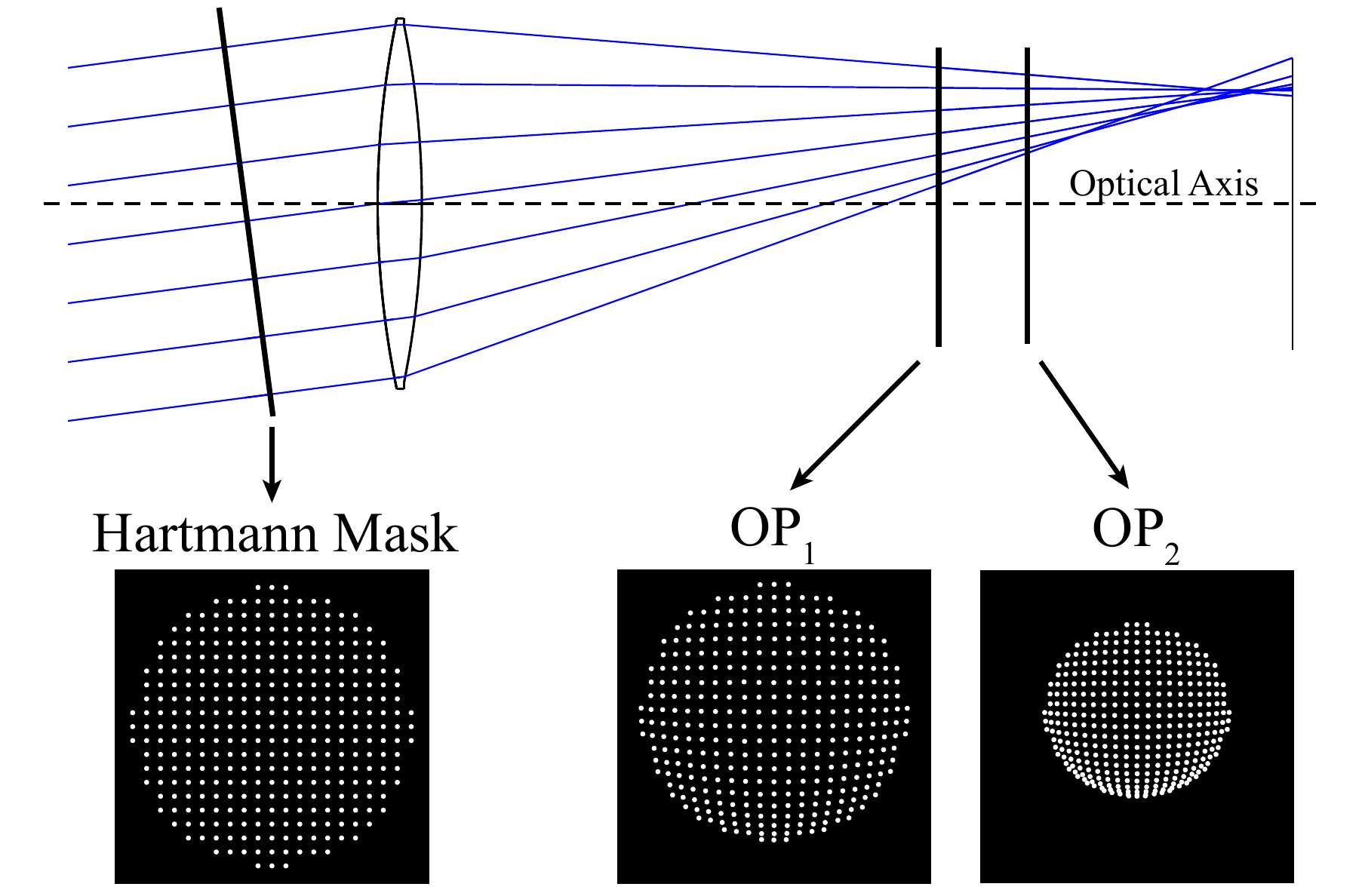}
        \caption{Our generalized Hartmann test}
        \label{fig:Hartmann_offaxis}
    \end{subfigure}
    \caption{Comparison between the classical Hartmann test (left), and our generalized Hartmann test (right).}
    \label{fig:Hartmann_compare}
\end{figure}

Figure \ref{fig:Hartmann_classic} shows the classical Hartmann test applied to a biconvex singlet lens. A mask with a known pattern of small holes, called the Hartmann mask, is placed at the aperture stop of the optical system. An image is then taken at a defocused plane $\mathrm{OP}_1$, called the observation plane \cite{HuertaCarranza2020}. In an ideal scenario in which there are no aberrations in the optical system, the reimaged pattern of holes at $\mathrm{OP}_1$ would be a scaled-down version of the pattern of holes on the Hartmann mask\cite{SalasPeimbert2005}. The deviations of the reimaged pattern of holes from an ideal pattern are called transverse ray aberrations. Wavefront error can be estimated from measurements of transverse ray aberrations.

\begin{figure}[H]
    \centering
    \begin{subfigure}{0.49\textwidth}
        \centering
        \includegraphics[width=0.9\textwidth]{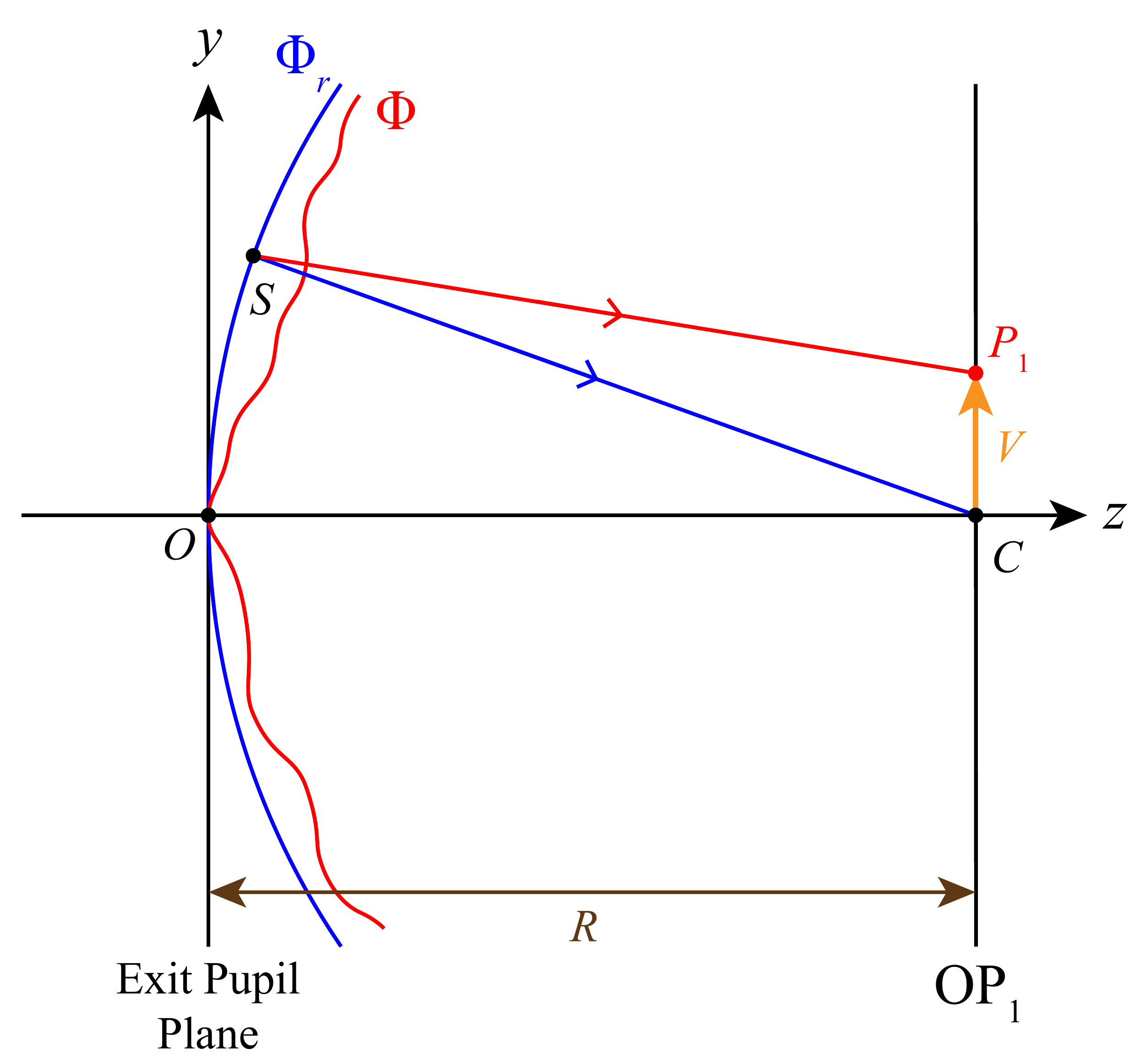}
        \caption{Classical Hartmann test with Rayces' equations}
        \label{fig:Hartmann_Rayces}
    \end{subfigure}
    \begin{subfigure}{0.49\textwidth}
        \centering
        \includegraphics[width=0.9\textwidth]{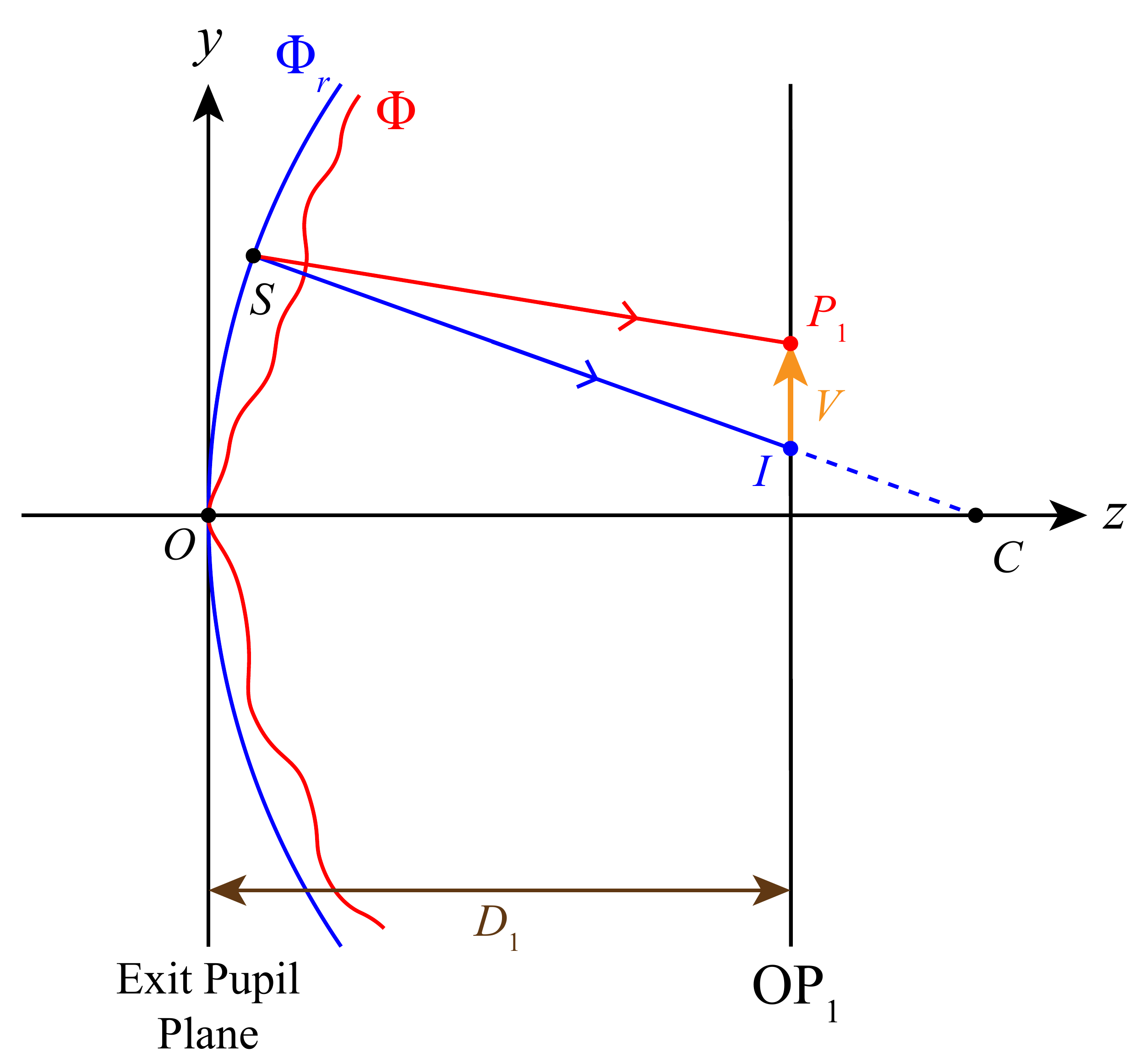}
        \caption{Classical Hartmann test in practice}
        \label{fig:Hartmann_Rayces_defocus}
    \end{subfigure}
    \caption{The geometry of the classical Hartmann test. The left figure shows the geometry of Rayces' equations \mbox{(Equation (\ref{eq:Rayces}))}, and the right figure shows the geometry of the classical Hartmann test applied in practice, with a defocused observation plane.}
    \label{fig:Hartmann_class}
\end{figure}

For a focal optical system, the reference wavefront $\Phi_r$ is usually taken to be spherical in shape, since an ideal lens should turn a plane wave into a spherical wave which focuses to a point. Let $R$ be the radius of the spherical reference wavefront. Then the relationship between the wavefront error $W$ and the transverse ray aberrations are given by a pair of nonlinear PDEs called Rayces' equations\cite{Rayces1964}:
\begin{align}\label{eq:Rayces}
    \frac{\partial W}{\partial x} &= -\frac{U}{R-W} \nonumber\\
    \frac{\partial W}{\partial y} &= -\frac{V}{R-W}
\end{align}
$U$ and $V$ are called the horizontal and vertical transverse ray aberrations respectively. The geometry of \mbox{Equation (\ref{eq:Rayces})} is illustrated in Figure \ref{fig:Hartmann_Rayces}. In Figure \ref{fig:Hartmann_Rayces}, the reference wavefront $\Phi_r$ is the blue curve and the actual wavefront $\Phi$ is the red curve. The exit pupil plane is at the origin $O$ and is perpendicular to the optical axis, which is the $+z$ axis. In other words, the exit pupil plane is the $z=0$ plane. The center of the spherical reference wavefront is at point $C$, which is the focus of the optical system. An ideal ray from point $S$ on $\Phi_r$ will pass through $C$. This is shown by the blue line $\overline{SC}$, which is perpendicular to $\Phi_r$. However, due to aberrations, a real ray from point $S$ will instead pass through a point $P_1$. This real ray, shown by the red line $\overline{SP_1}$, is perpendicular to $\Phi$. The vertical transverse ray aberration $V$ is shown by the orange vector, and is equal to the vertical distance between $P_1$ and $C$. Here, the observation plane $\mathrm{OP}_1$ is at the focal plane.

With measurements of $U$ and $V$, $W$ can be estimated using Equation (\ref{eq:Rayces}). However, in practice, \mbox{Equation (\ref{eq:Rayces})} is linearized so that it is easier to solve; we assume that $R \gg W$, so that we can write \cite{MalacaraOST,Rayces1964,HuertaCarranza2020,MalacaraHernndez2015,SalasPeimbert2005}:
\begin{align}\label{eq:Rayces_approx}
    \frac{\partial W}{\partial x} \approx -\frac{U}{R} \nonumber\\
    \frac{\partial W}{\partial y} \approx -\frac{V}{R}
\end{align}
Equation (\ref{eq:Rayces_approx}) is used in practice. It is crucial to note that $x$ and $y$ in Equations (\ref{eq:Rayces}) and (\ref{eq:Rayces_approx}) refer to points of intersection between real rays and the reference wavefront\cite{Welford1986}, i.e. the points $S$. $x$ and $y$ in these equations do not refer to points of intersection between real rays and the exit pupil plane.

Note that the Hartmann test is a geometrical optics technique which relies on ray tracing. While Rayces' equations works if $\mathrm{OP}_1$ at the focal plane, this cannot be done in practice because we cannot adequately identify the reimaged holes at the focal plane. In practice, $\mathrm{OP}_1$ must be outside of the caustic zone, where the rays do not cross \cite{MalacaraOST,SalasPeimbert2005}. $\mathrm{OP}_1$ must be a defocused plane, and this is illustrated in Figure \ref{fig:Hartmann_Rayces_defocus}. In Figure \ref{fig:Hartmann_Rayces_defocus}, $\mathrm{OP}_1$ is at a distance of $D_1$ from the exit pupil plane, where $D_1 < R$. An ideal ray from point $S$ on $\Phi_r$ now intersects $\mathrm{OP}_1$ at point $I$. The vertical transverse ray aberration $V$ is the vertical distance between $P_1$ and $I$. We can see that the $V$ in Figure \ref{fig:Hartmann_Rayces_defocus} is not the same as the $V$ in Figure \ref{fig:Hartmann_Rayces}. That is, the $V$ in Figure \ref{fig:Hartmann_Rayces_defocus} is not the same as the $V$ in Equations (\ref{eq:Rayces}) and (\ref{eq:Rayces_approx}). Transverse ray aberrations measured on a defocused plane cannot immediately be used with Equations (\ref{eq:Rayces}) and (\ref{eq:Rayces_approx}).

In practice, there are several approaches to account for the fact that we measure transverse ray aberrations at a defocused plane rather than at the focal plane. One approach is to use a modified version of Rayces' equations containing a multiplicative factor which accounts for transverse ray aberrations measured on a defocused plane\cite{TllezQuiones2016}. Another more common approach is to scale down the Hartmann mask hole pattern such that the separations between the reimaged holes on $\mathrm{OP}_1$ and the corresponding scaled-down Hartmann mask holes are minimized; transverse ray aberrations are then taken to be the differences between the reimaged hole positions and the scaled-down Hartmann mask hole positions\cite{MalacaraOST,SalasPeimbert2005}, and these transverse ray aberrations are then used with Equation (\ref{eq:Rayces_approx}). However, in this approach, there is an implicit assumption that $x$ and $y$ correspond to the positions of the holes on the Hartmann mask. This implies that $x$ and $y$ refer to the points of intersection between real rays and the exit pupil plane, instead of the reference wavefront, which is incorrect\cite{Welford1986}. This may be sufficiently true for rays with small angles with respect to the optical axis ($+z$ axis), but this is not true in general. As a result of this, and other reasons\cite{Mejia2012}, Equation (\ref{eq:Rayces_approx}) can fail for small $f$/\# (fast beams).

In these approaches, there is also an implicit assumption that the reference wavefront is a sphere that corresponds to an on-axis field. As illustrated in Figure \ref{fig:Hartmann_class}, the center $C$ of the spherical reference wavefront is on the optical axis. That is:
\begin{equation}
    \Phi_r(x,y) = -\sqrt{R^2 - x^2 - y^2} + R
\end{equation}
However, for G-CLEF's $f$/2.25 red camera, we would like to estimate the wavefront error for off-axis field angles; that is, we would like the reference wavefront to be a sphere with a center not on the optical axis. This, along with the fact that Equation (\ref{eq:Rayces_approx}) can fail for small $f$/\#, motivates the need to develop a more generalized method. In Sections \ref{sec:overview}, \ref{sec:TRA}, and \ref{sec:TRA_to_WFE}, we discuss a generalization of the classical Hartmann test for off-axis field angles. Our method differs from the classical Hartmann test in that we take images on two defocused observation planes to trace rays back to the reference wavefront, so that the points of intersection ($S$) with the reference wavefront can be accurately determined. This is illustrated in Figure \ref{fig:Hartmann_offaxis}.

%% file: Sections/app.tex
\section{Hartmann Test Apparatus for the Red Camera Test Setup}\label{sec:Hartmann_app}

\begin{figure}[H]
  \centering
  \includegraphics[width=0.8\textwidth]{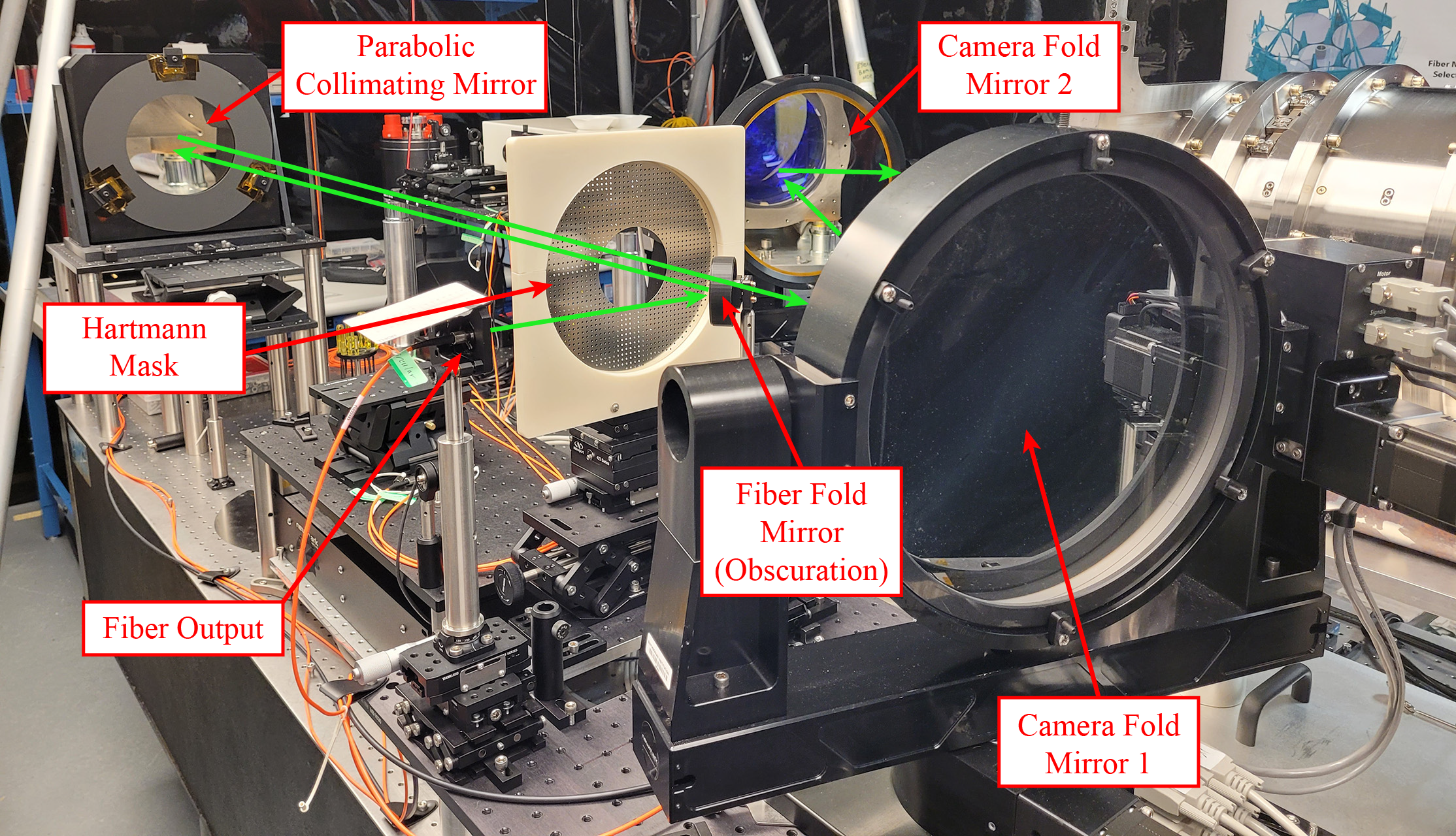}
  \caption{Real-world photo of the test setup in Figure \ref{fig:Red_AIT_m114}, with our Hartmann mask in place. We placed the Hartmann mask in between the parabolic collimating mirror and the fiber fold mirror, perpendicular to the optical axis. We placed the mask on a series of stages, and adjusted it such that the chief ray passes through the center of the mask.}
  \label{fig:Red_AIT_lab_Hartmann}
\end{figure}

To use the Hartmann test with our red camera test setup discussed in Section \ref{sec:red_cam_test}, we laser-cut a custom Hartmann mask and placed it between the parabolic mirror and the fiber fold mirror, perpendicular to the optical axis. Our Hartmann mask was made of 0.010 inch thick 302/304 full hard cross stainless steel, primed with Aeroglaze 9929, and coated with Aeroglaze Z306 on one side. Our mask was held by a 3D-printed mount, which we placed on three translating stages. We positioned our mask such that the chief ray of the system passes through the center of the mask. Figure \ref{fig:Red_AIT_lab_Hartmann} shows the Hartmann mask in our test setup. The position of the Hartmann mask is also labeled in Figure \ref{fig:Red_AIT_m114}.

\begin{figure}[H]
    \centering
    \begin{subfigure}{0.32\textwidth}
        \centering
        \includegraphics[width=\textwidth]{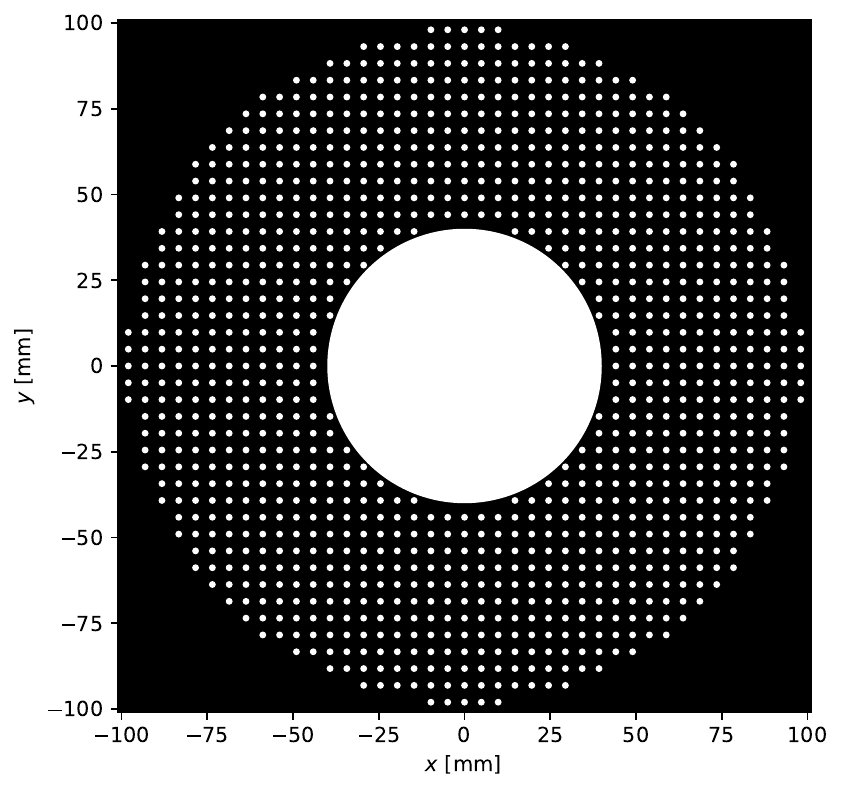}
        \caption{Our Hartmann mask}
        \label{fig:Hartmann_mask}
    \end{subfigure}
    \begin{subfigure}{0.32\textwidth}
        \centering
        \includegraphics[width=\textwidth]{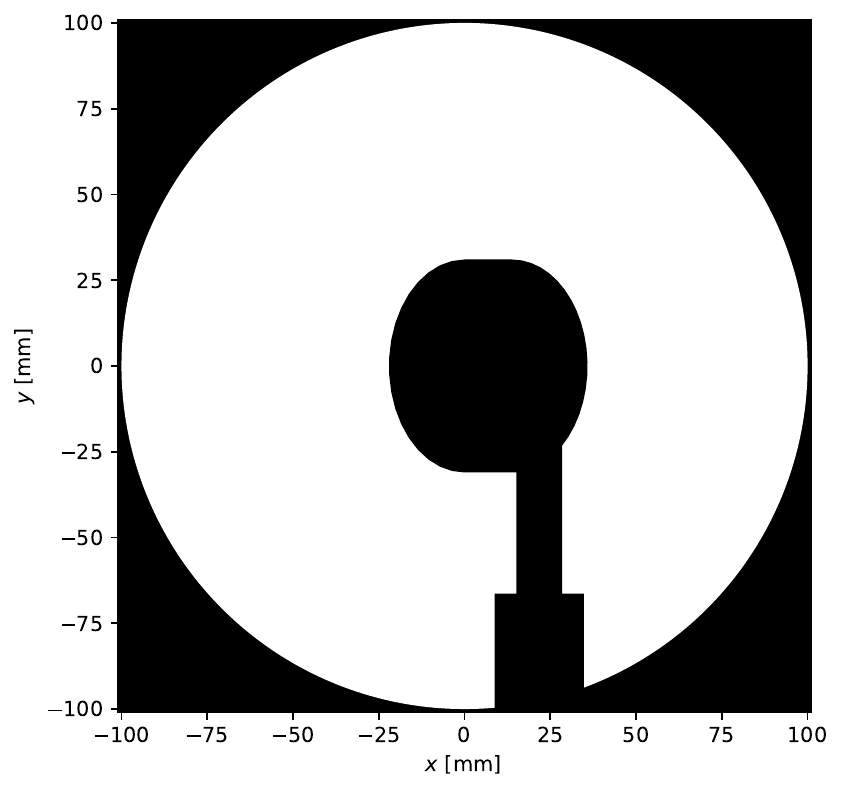}
        \caption{Fiber fold mirror obscuration}
        \label{fig:obs}
    \end{subfigure}
    \begin{subfigure}{0.32\textwidth}
        \centering
        \includegraphics[width=\textwidth]{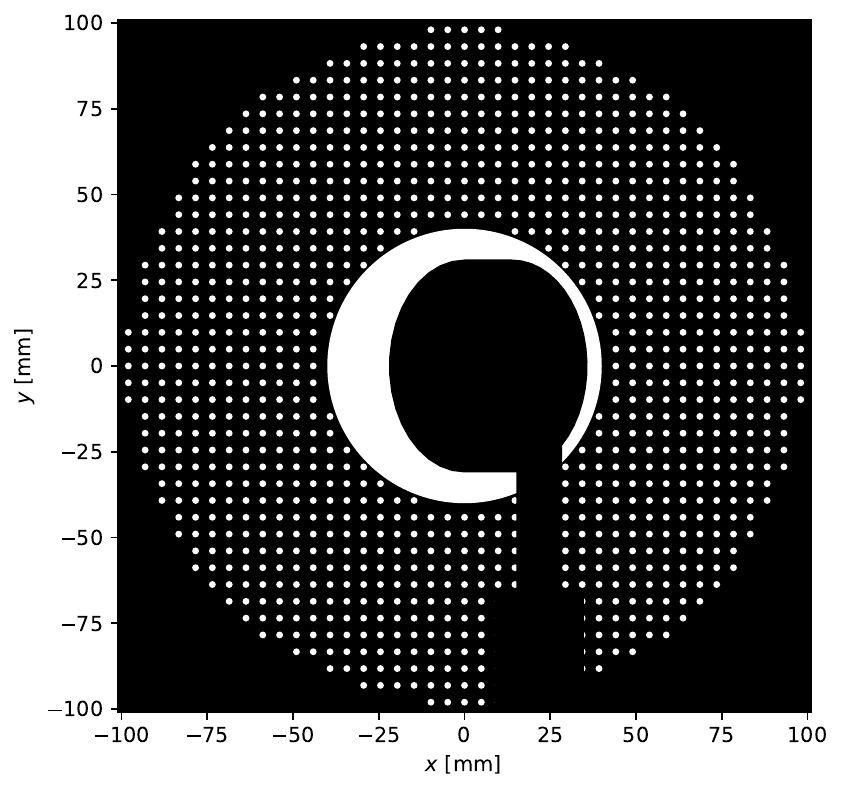}
        \caption{Mask and obscuration combined}
        \label{fig:Hartmann+obs}
    \end{subfigure}
    \caption{Our Hartmann mask (left), the fiber fold mirror obscuration (center), and the resulting effective aperture from the mask and obscuration (right).}
\end{figure}

The design of our Hartmann mask is shown in Figure \ref{fig:Hartmann_mask}. Our mask consists of a grid of small holes, each with a radius of 1 mm, with centers separated by 4.9 mm. The extent of our mask is 200 mm, which is the diameter of beam collimated by the parabolic mirror. At the center of our mask is a larger hole with a radius of 40 mm. This hole allows the diverging beam reflected by the fiber fold mirror (traveling towards the left in \mbox{Figure \ref{fig:Red_AIT_m114}}) to make a first pass through the mask unobstructed. The diverging beam is reflected and collimated by the parabolic mirror and makes a second pass through the Hartmann mask (as it travels towards the right in \mbox{Figure \ref{fig:Red_AIT_m114}}). The collimated beam then passes through the fiber fold mirror, which is an obscuration (see \mbox{Figure \ref{fig:obs}}). The Hartmann mask and fiber fold mirror obscuration together result in an effective aperture shown in \mbox{Figure \ref{fig:Hartmann+obs}}. This pattern of collimated light is then injected into the red camera at an off-axis field angle by the two camera fold mirrors.

%% file: Sections/method_overview.tex
\section{Overview of our Method}\label{sec:overview}

In this section, we discuss an overview of our generalized method, which consists of two parts: measuring transverse ray aberrations (\mbox{Section \ref{sec:TRA}}), and estimating the wavefront error from transverse ray aberration measurements (Section \ref{sec:TRA_to_WFE}). Our method differs from the classical Hartmann test in that we take images at two defocused observation planes and we consider off-axis field angles (see Figure \ref{fig:Hartmann_compare}). In Sections  \ref{sec:TRA} and \ref{sec:TRA_to_WFE}, we use a specific example to demonstrate our method; we use the example of the two camera fold mirrors injecting collimated light into the red camera with ray trajectories corresponding to G-CLEF's $m=114$ diffraction order at $\lambda=536$ nm, as in Figures \ref{fig:Red_AIT_m114} and \ref{fig:Red_AIT_standalone}.


\subsection{Geometry of our Method}

\begin{figure}[htb!]
  \centering
  \includegraphics[width=0.65\textwidth]{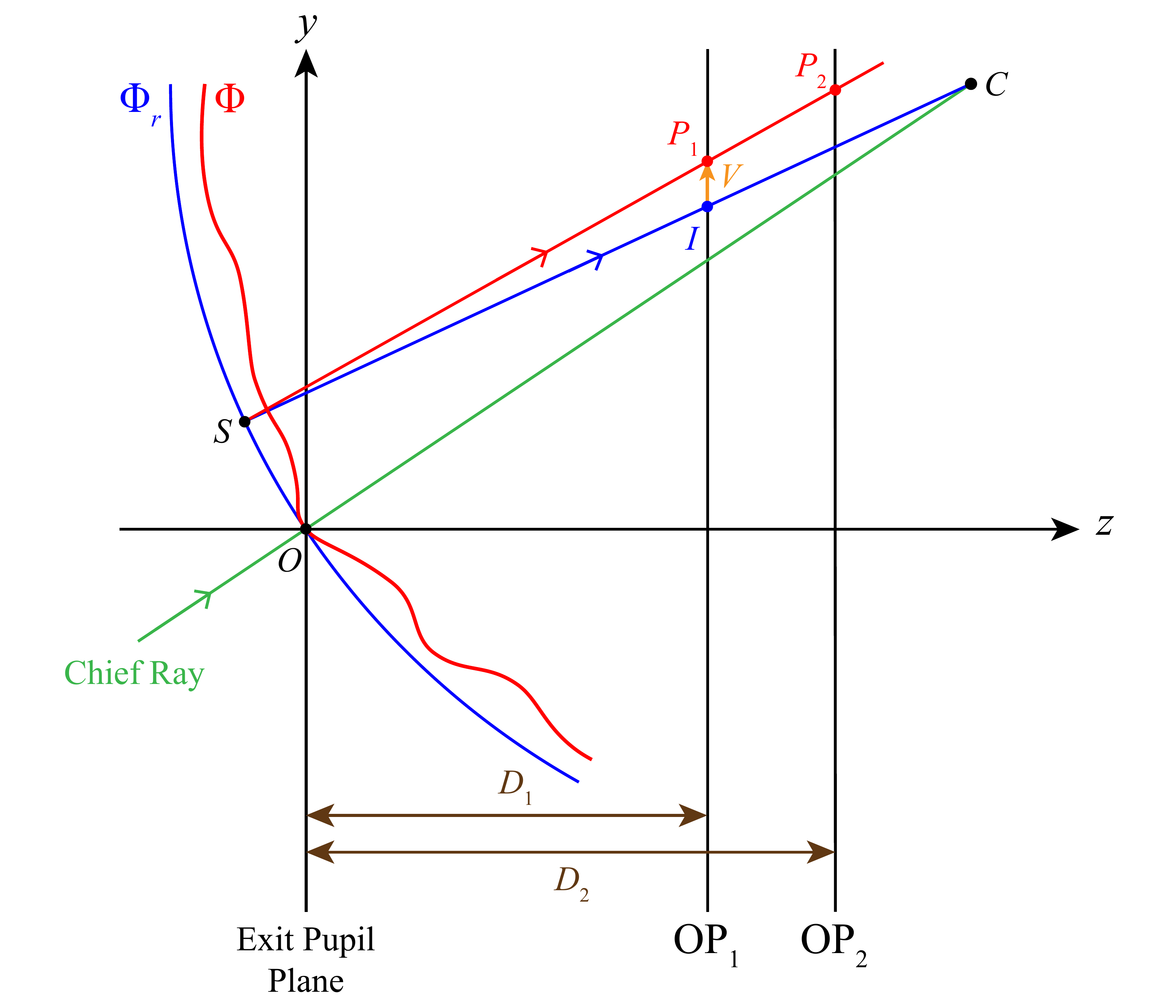}
  \caption{The geometry of our generalized Hartmann test. We take images at two defocused planes $\mathrm{OP}_1$ and $\mathrm{OP}_2$. The reference wavefront $\Phi_r$ is a sphere with its center at $C$, and the chief ray is the green line. The actual wavefront is $\Phi$. For the real ray $\overline{SP_1}$ and the ideal ray $\overline{SI}$, the vertical transverse ray aberration on $\mathrm{OP}_1$ is $V$.}
  \label{fig:TRA_WFE}
\end{figure}

Figure \ref{fig:TRA_WFE} illustrates the geometry of our generalized method. As in Figure \ref{fig:Hartmann_class}, the origin is point $O$, and the exit pupil plane is the $z=0$ plane, which is perpendicular to the optical axis (the $+z$ axis). The reference wavefront $\Phi_r$ is the blue curve, which is a sphere centered at point $C$. $C$ is the focus of the optical system for an off-axis field with a chief ray shown in green. The chief ray is coincident with the line segment $\overline{OC}$. The length of $\overline{OC}$ is $R$, which is the radius of the spherical reference wavefront. $\Phi_r$ and the chief ray both cross the optical axis at $O$. The actual wavefront $\Phi$ is the red curve.

We take images on two defocused observation planes $\mathrm{OP}_1$ and $\mathrm{OP}_2$, which are at distances of $D_1$ and $D_2$ respectively from the exit pupil plane. $\mathrm{OP}_1$ and $\mathrm{OP}_2$ are perpendicular to the optical axis. A real ray, shown by the red line, intersects $\mathrm{OP}_1$ and $\mathrm{OP}_2$ at points $P_1$ and $P_2$ respectively. With knowledge of $P_1$ and $P_2$, we can trace the real ray back to $\Phi_r$ and determine the point $S$, which is the point of intersection between the real ray and $\Phi_r$. Note that the real ray is perpendicular to $\Phi$. An ideal ray from point $S$ on $\Phi_r$, shown by the blue line, passes through point $C$, intersecting $\mathrm{OP}_1$ at point $I$. The vertical distance between $P_1$ and $I$ is the vertical transverse ray aberration $V$. We would like to measure $V$, and then use that to estimate the wavefront error $W\equiv \Phi-\Phi_r$.

\subsection{Summary of our Method}

Our method can be summarized in four steps:
\begin{enumerate}
    \item Consider one real ray (red line in Figure \ref{fig:TRA_WFE}). We first measure the positions of the points $P_1$ and $P_2$ on $\mathrm{OP}_1$ and $\mathrm{OP}_2$ (Section \ref{sec:TRA_centroids}).
    \item Using $P_1$ and $P_2$, we trace the real ray back to point $S$ on $\Phi_r$ (Section \ref{sec:TRA_trace_back}).
    \item From point $S$, we propagate an ideal ray 
    (blue line in Figure \ref{fig:TRA_WFE}) forward to $\mathrm{OP}_1$, to point $I$. We can then calculate the transverse ray aberrations on $\mathrm{OP}_1$ (Section \ref{sec:TRA_prop_forward}).
    \item With the transverse ray aberration measurements, we can estimate the wavefront error using a pair of nonlinear PDEs (Equation (\ref{eq:TRA_WFE_PDE}) in Section \ref{sec:TRA_to_WFE}).
\end{enumerate}

%% file: Sections/TRA.tex
\section{Part I: Measuring Transverse Ray Aberrations}\label{sec:TRA}


\subsection{Reference Wavefront}

From the geometry in Figure \ref{fig:TRA_WFE}, we can see that the equation for the spherical reference wavefront is:
\begin{equation}\label{eq:phi_r}
    \Phi_r(x,y) \equiv -\sqrt{R^2 - (x-c_x)^2 - (y-c_y)^2} + \sqrt{R^2 - c_x^2 - c_y^2}
\end{equation}
where the center $C$ of the sphere is $\left(c_x,c_y,\sqrt{R^2 - c_x^2 - c_y^2}\right)$. The partial derivatives of Equation (\ref{eq:phi_r}) are:
\begin{align}
    \frac{\partial \Phi_r}{\partial x} &= \frac{(x-c_x)}{\sqrt{R^2 - (x-c_x)^2 - (y-c_y)^2}} \nonumber \\
    \frac{\partial \Phi_r}{\partial y} &= \frac{(y-c_y)}{\sqrt{R^2 - (x-c_x)^2 - (y-c_y)^2}} \label{eq:phi_r_deriv}
\end{align}


\subsection{Identify Centroids on Observation Planes}\label{sec:TRA_centroids}

\begin{figure}[H]
    \centering
    \begin{subfigure}{0.49\textwidth}
        \centering
        \includegraphics[width=0.85\textwidth]{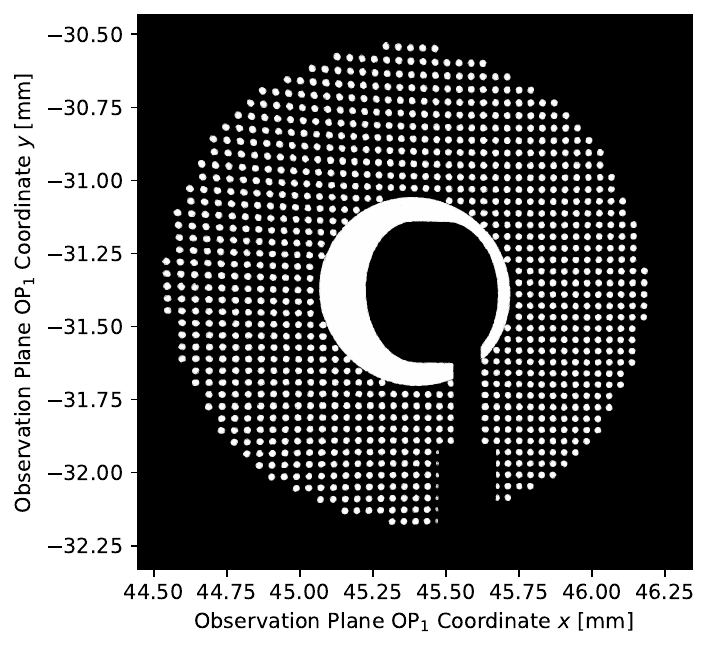}
        \caption{$\mathrm{OP}_1$ simulated image}
        \label{fig:OP1}
    \end{subfigure}
    \begin{subfigure}{0.49\textwidth}
        \centering
        \includegraphics[width=0.85\textwidth]{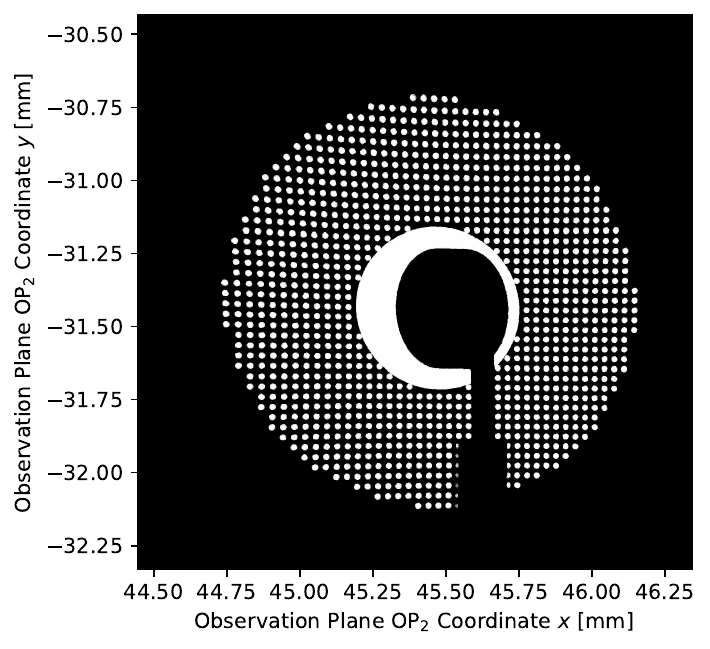}
        \caption{$\mathrm{OP}_2$ simulated image}
        \label{fig:OP2}
    \end{subfigure}
    \caption{Simulated observation plane images from Zemax OpticStudio, for the setup in Figure \ref{fig:Red_AIT_m114}, where rays are injected into the G-CLEF red camera at trajectories corresponding to G-CLEF's $m=114$ diffraction order at $\lambda=536$ nm. We used Zemax's Geometric Image Analysis feature to simulate these images. For each image, around one million rays were launched into the system. Here, $\mathrm{OP}_1$ was 3.5 mm before focus, and $\mathrm{OP}_2$ was 3.0 mm before focus. Compare these images with Figure \ref{fig:Hartmann+obs}.}
    \label{fig:OP}
\end{figure}

Figure \ref{fig:OP} shows images of the observation planes $\mathrm{OP}_1$ and $\mathrm{OP}_2$, simulated using Zemax OpticStudio. Here, $\mathrm{OP}_1$ was 3.5 mm before focus, and $\mathrm{OP}_2$ was 3.0 mm before focus. Since the observation planes are before focus, where the beam is converging, the extent of $\mathrm{OP}_1$ is larger than the extent of $\mathrm{OP}_2$. We need to identify the centroids of the reimaged holes on the observation planes, and then match corresponding pairs of centroids in $\mathrm{OP}_1$ and $\mathrm{OP}_2$. In other words, we want to find corresponding pairs of points $P_1$ and $P_2$ (see Figure \ref{fig:TRA_WFE}). To identify individual reimaged holes in each image, we first binary thresholded the image and rejected the reimaged holes that were partially obscured by the fiber fold mirror. We then found connected components in the thresholded image, using 4-connectivity\cite{Bolelli2022,opencv_library}. Each connected component consists of pixels belonging to a particular reimaged hole. We then computed the centroid of each connected component. Next, we noticed that the pattern in $\mathrm{OP}_2$ is nearly a scaled-down version of the pattern in $\mathrm{OP}_1$. Hence, to match centroids in the two observation planes, we scaled down the centroids in $\mathrm{OP}_1$ and compared this with the centroids in $\mathrm{OP}_2$. For each centroid in $\mathrm{OP}_2$, we found the nearest centroid in the scaled-down version of $\mathrm{OP}_1$, using a $k$-d tree \cite{Bentley1975,scikit-learn}. Then each centroid in $\mathrm{OP}_2$ was matched to its corresponding centroid in the original $\mathrm{OP}_1$. The centroids are shown in Figure \ref{fig:OP_compare}. In this plot, the green points are the centroids for $\mathrm{OP}_1$ (the points $P_1$) and the orange points are the centroids for $\mathrm{OP}_2$ (the points $P_2$).

\subsection{Trace Real Rays Backward to Reference Wavefront}\label{sec:TRA_trace_back}

\begin{figure}[H]
    \centering
    \begin{subfigure}{0.5\textwidth}
        \centering
        \includegraphics[width=\textwidth]{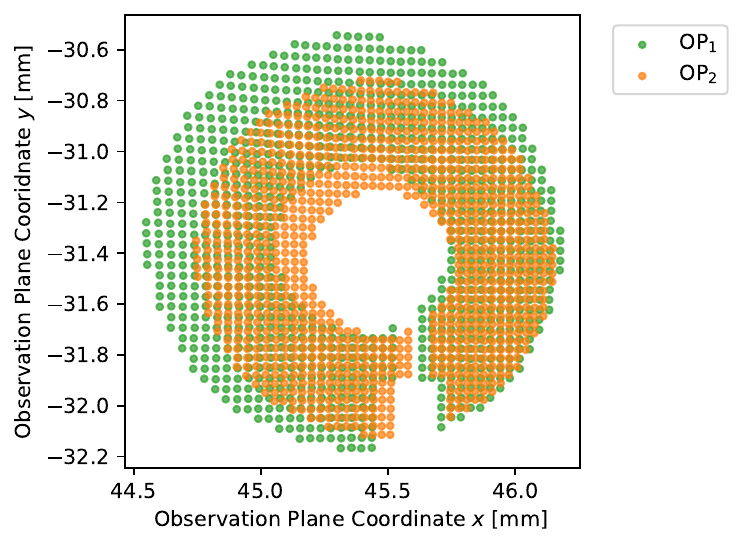}
        \caption{Identified centroids in Figure \ref{fig:OP}. The green points correspond to the points $P_1$ and the orange points correspond to the points $P_2$.}
        \label{fig:OP_compare}
    \end{subfigure}
    \hfill
    \begin{subfigure}{0.38\textwidth}
        \centering
        \includegraphics[width=\textwidth]{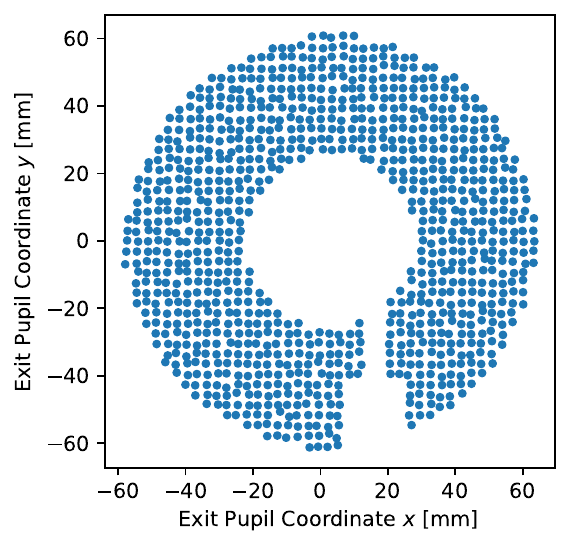}
        \caption{Points of intersection (points $S$) between real rays and $\Phi_r$, computed using corresponding pairs of $P_1$ and $P_2$ in Figure \ref{fig:OP_compare}.}
        \label{fig:POI}
    \end{subfigure}
    \caption{The points $P_1$ and $P_2$ (left), and the points $S$ (right).}
\end{figure}

Next, we traced each pair of points $P_1$ and $P_2$ back to the reference wavefront $\Phi_r$ in order to obtain the \mbox{points $S$} (see Figure \ref{fig:TRA_WFE}). Each pair of points $P_1$ and $P_2$ corresponds to a real ray, and so we can express a real ray as a line in $\mathbb{R}^3$. With two points $P_1$ and $P_2$, we can obtain a parametric equation for the line. Tracing a ray back to $\Phi_r$ merely means finding the point of intersection between a line and a sphere. We describe how this is done in \mbox{Appendix \ref{sec:app_line_sphere}}. The points of intersection are shown in Figure \ref{fig:POI}. 

\subsection{Propagate Ideal Rays Forward to Observation Plane}\label{sec:TRA_prop_forward}

With the points $S$, the next step is to propagate ideal rays forward towards one of the two observation planes, so that we can compute transverse ray aberrations on that observation plane. Following Figure \ref{fig:TRA_WFE}, let us propagate ideal rays towards $\mathrm{OP}_1$. Consider one ideal ray. Its vertical height at the point of intersection ($S$) is $y$. Since the ideal ray is perpendicular to $\Phi_r$, it follows that the derivative of the ideal ray's vertical height with respect to $+z$ is $-\frac{\partial \Phi_r}{\partial y}$. From the reference wavefront to $\mathrm{OP}_1$, this ideal ray travels a distance of $(D_1-\Phi_r)$ in the $+z$ direction. Hence, the vertical height of this ideal ray at $\mathrm{OP}_1$ is:
\begin{equation}
    \tilde{y}_1 = y + \left(-\frac{\partial \Phi_r}{\partial y} \right) (D_1 - \Phi_r) \label{eq:tilde_y1}
\end{equation}
Likewise, the horizontal height of this ideal ray at $\mathrm{OP}_1$ is:
\begin{equation}
    \tilde{x}_1 = x + \left(-\frac{\partial \Phi_r}{\partial x} \right) (D_1 - \Phi_r) \label{eq:tilde_x1}
\end{equation}
The point $I$ in Figure \ref{fig:TRA_WFE} has \mbox{coordinates $(\tilde{x}_1,\tilde{y}_1,D_1)$}.

\begin{figure}[H]
  \centering
  \includegraphics[width=0.55\textwidth]{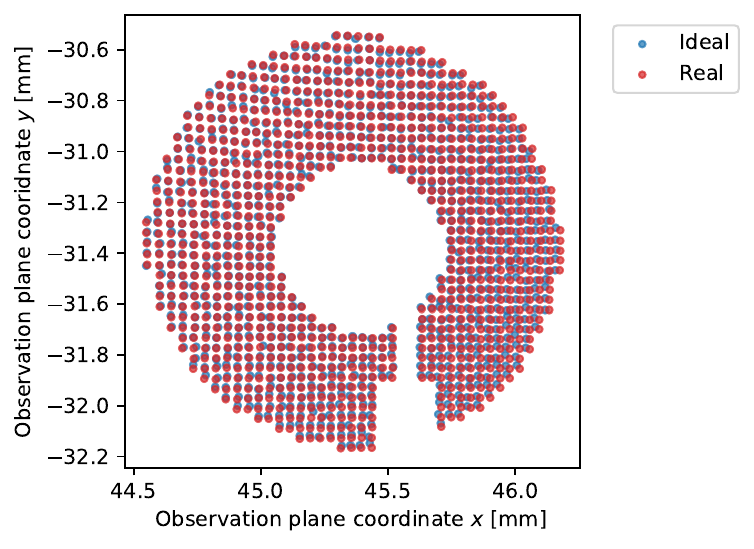}
  \caption{Comparison between the positions of real rays (red points) and ideal rays (blue points) on $\mathrm{OP}_1$. The real ray positions correspond to points $P_1$ and the ideal ray positions correspond to points $I$. The red points here are the same as the green points in Figure \ref{fig:OP_compare}. The positions of the blue points here are computed using Equations (\ref{eq:tilde_y1}) and (\ref{eq:tilde_x1}).}
  \label{fig:OP_real_ideal}
\end{figure}

\begin{figure}[H]
    \centering
    \begin{subfigure}{0.49\textwidth}
        \centering
        \includegraphics[width=\textwidth]{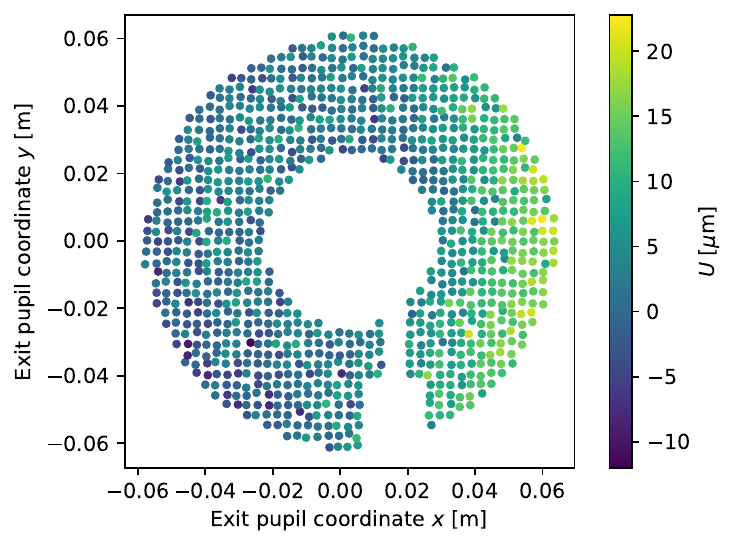}
        \caption{Horizontal transverse ray aberration $U$}
        \label{fig:TRA_U}
    \end{subfigure}
    \begin{subfigure}{0.49\textwidth}
        \centering
        \includegraphics[width=\textwidth]{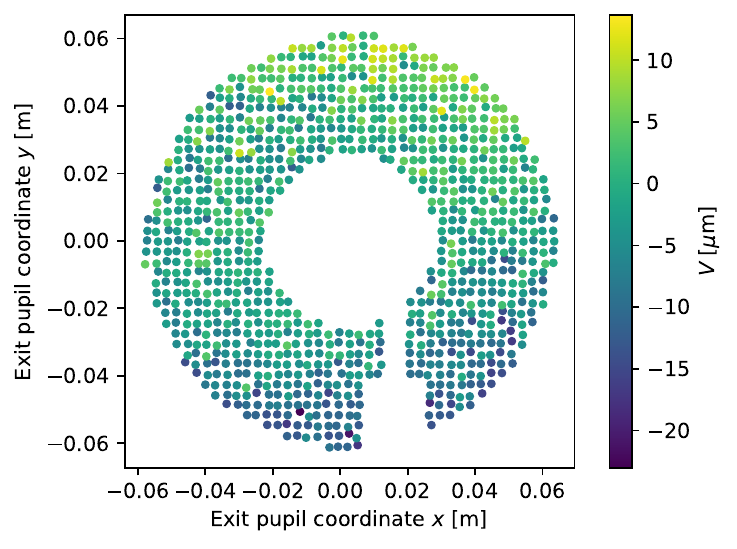}
        \caption{Vertical transverse ray aberration $V$}
        \label{fig:TRA_V}
    \end{subfigure}
    \caption{Horizontal and vertical transverse ray aberrations computed using the positions in Figure \ref{fig:OP_real_ideal}.}
    \label{fig:TRA}
\end{figure}

Now, let the vertical and horizontal height of the corresponding real ray at $\mathrm{OP}_1$ be $\tilde{y}_1'$ and $\tilde{x}_1'$ respectively. That is, $(\tilde{x}_1',\tilde{y}_1',D_1)$ are the coordinates for the corresponding point $P_1$. Then the vertical and horizontal transverse ray aberrations are given by Equations (\ref{eq:V}) and (\ref{eq:U}) respectively:
\begin{align}
    V &\equiv \tilde{y}_1' - \tilde{y}_1 \label{eq:V} \\
    U &\equiv \tilde{x}_1' - \tilde{x}_1 \label{eq:U}
\end{align}
The real and ideal ray positions on $\mathrm{OP}_1$ are shown in Figure \ref{fig:OP_real_ideal}. The horizontal and vertical transverse ray aberrations are shown in Figure \ref{fig:TRA}.

%% file: Sections/TRA_to_WFE.tex
\section{Part II: Transverse Ray Aberrations to Wavefront Error}\label{sec:TRA_to_WFE}

\subsection{Derivation of PDEs Relating Transverse Ray Aberrations to Wavefront Error}

Now, we would like to relate transverse ray aberrations to wavefront error. To do this, first, let us recall the equation for propagating an ideal ray from the point of intersection forward towards $\mathrm{OP}_1$, which is Equation (\ref{eq:tilde_y1}). We can replace $\Phi_r$ in Equation (\ref{eq:tilde_y1}) with $\Phi$ to obtain the vertical height of the corresponding real ray at $\mathrm{OP}_1$:
\begin{equation} \label{eq:tilde_y1p}
    \tilde{y}_1' = y + \left(-\frac{\partial \Phi}{\partial y} \right) (D_1 - \Phi)
\end{equation}
Then by substituting Equations (\ref{eq:tilde_y1}) and (\ref{eq:tilde_y1p}) into Equation (\ref{eq:V}), we can obtain a nonlinear PDE relating the wavefront error $W$ to the vertical transverse ray aberration $V$:
\begin{align*}
    V &\equiv \tilde{y}_1' - \tilde{y}_1 && \textrm{Start with Equation (\ref{eq:V})}\\
     &= \left[y + \left(-\frac{\partial \Phi}{\partial y} \right) (D_1 - \Phi)\right] - \left[y + \left(-\frac{\partial \Phi_r}{\partial y} \right) (D_1 - \Phi_r)\right] && \textrm{Substitute in Equations (\ref{eq:tilde_y1}) and (\ref{eq:tilde_y1p})}\\
    &= \left(-\frac{\partial \Phi}{\partial y} \right) (D_1 - \Phi) - \left(-\frac{\partial \Phi_r}{\partial y} \right) (D_1 - \Phi_r) \\
    &= \left(-\frac{\partial (W + \Phi_r)}{\partial y} \right) (D_1 - (W + \Phi_r)) - \left(-\frac{\partial \Phi_r}{\partial y} \right) (D_1 - \Phi_r) && \textrm{Substitute in Equation (\ref{eq:WFE})}\\
    &= \left(-\frac{\partial W}{\partial y} - \frac{\partial \Phi_r}{\partial y}\right) (D_1 - W - \Phi_r) + (D_1 - \Phi_r) \frac{\partial \Phi_r}{\partial y} \\
    &= -(D_1-W-\Phi_r)\frac{\partial W}{\partial y}  + W\frac{\partial \Phi_r}{\partial y}
\end{align*}
We can use the same reasoning with the horizontal transverse ray aberration $U$. We then obtain a pair of nonlinear PDEs relating transverse ray aberrations to wavefront error:
\begin{align}
    U &= -(D_1-W-\Phi_r)\frac{\partial W}{\partial x}  + W\frac{\partial \Phi_r}{\partial x} \nonumber \\
    V &= -(D_1-W-\Phi_r)\frac{\partial W}{\partial y}  + W\frac{\partial \Phi_r}{\partial y} \label{eq:TRA_WFE_PDE}
\end{align}
Note that in these PDEs, the reference wavefront $\Phi_r$ does not necessarily have to have the same form as in Equation (\ref{eq:phi_r}). These PDEs allow for an arbitrary reference wavefront, as long as $x$ and $y$ correspond to the points of intersection between real rays and the reference wavefront.


\subsection{Solving the PDEs}\label{sec:solve}

To solve Equation (\ref{eq:TRA_WFE_PDE}), we used a modal approach. We expressed the wavefront error $W$ in Equation (\ref{eq:TRA_WFE_PDE}) as a linear combination of Zernike circle polynomials:
\begin{equation}\label{eq:W_Zern}
    W(x,y) = \sum_{j=1}^N A_j Z_j\left(\frac{x}{R_\mathrm{EXPR}},\frac{y}{R_\mathrm{EXPR}}\right)
\end{equation}
$Z_j$ is the $j$th Zernike polynomial according to the Noll indexing scheme\cite{Noll1976} and $A_j$ is the Zernike mode amplitude for the $j$th Zernike polynomial. See Appendix \ref{sec:app_Zernike} for a discussion on the Zernike polynomials we used. $N$ is a positive integer which represents the number of Zernike polynomials used to decompose $W$. We selected $N=22$. Since the Zernike polynomials are only defined on the unit disk, we normalized $x$ and $y$ with respect to the exit pupil radius $R_\mathrm{EXPR}$. With this Zernike decomposition of $W$, we can write the partial derivatives of $W$ as:
\begin{align}
    \frac{\partial}{\partial x} W(x,y) &= \sum_{j=1}^N A_j \frac{\partial}{\partial x} \left[Z_j\left(\frac{x}{R_\mathrm{EXPR}},\frac{y}{R_\mathrm{EXPR}}\right]\right) \label{eq:dW_dx}\\
    \frac{\partial}{\partial y} W(x,y) &= \sum_{j=1}^N A_j \frac{\partial}{\partial y} \left[Z_j\left(\frac{x}{R_\mathrm{EXPR}},\frac{y}{R_\mathrm{EXPR}}\right]\right) \label{eq:dW_dy}
\end{align}
Now, note that the partial derivatives of the Zernike polynomials in Cartesian coordinates can also be expressed as a linear combination of Zernike polynomials\cite{Sch2016,Stephenson2014}. These are given by Equations (\ref{eq:dZ_dx}) and (\ref{eq:dZ_dy}) in \mbox{Appendix \ref{sec:app_Zernike}}.

In order to solve the nonlinear PDEs, we substituted Equations (\ref{eq:dZ_dx}) and (\ref{eq:dZ_dy}) into Equations (\ref{eq:dW_dx}) and (\ref{eq:dW_dy}) respectively, and then we substituted Equations (\ref{eq:W_Zern}), (\ref{eq:dW_dx}), and (\ref{eq:dW_dy}) into Equation (\ref{eq:TRA_WFE_PDE}). We then used nonlinear least squares regression\cite{2020SciPy-NMeth} to simultaneously fit $A_1,\ldots,A_N$ in the two nonlinear PDEs. We discuss the results in Section \ref{sec:results}.

%% file: Sections/results.tex
\section{Results and Discussion}\label{sec:results}

\begin{figure}[H]
  \centering
  \includegraphics[width=0.65\textwidth]{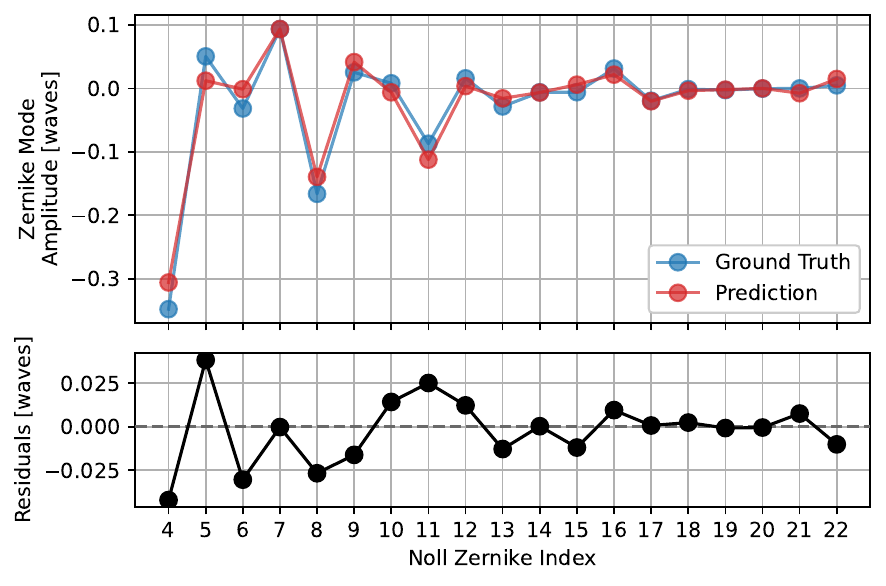}
  \caption{Wavefront error in the G-CLEF red camera predicted by our method, when applied to simulated observation plane images for the setup in Figure \ref{fig:Red_AIT_m114}, where rays are injected into the red camera at trajectories corresponding to G-CLEF's $m=114$ diffraction order at $\lambda=536$ nm. The blue points are the ground truth Zernike mode amplitudes for the wavefront error from Zemax OpticStudio. The red points are the Zernike mode amplitudes for the wavefront error predicted by our method, obtained by solving Equation (\ref{eq:TRA_WFE_PDE}) with the transverse ray aberrations in Figure \ref{fig:TRA} and the points of intersection in Figure \ref{fig:POI}.}
  \label{fig:zernike_results}
\end{figure}

Figure \ref{fig:zernike_results} shows the results of our method applied to simulated images from G-CLEF's red camera, where rays are injected into the red camera at trajectories corresponding to G-CLEF's $m=114$ diffraction order at \mbox{$\lambda=536$ nm}. The blue points are the ground truth Zernike decomposition of the wavefront error from Zemax OpticStudio, and the red points are the Zernike mode amplitudes of the wavefront error predicted by our method. Piston and tilt Zernike mode amplitudes ($j=1,2,3$) are not reported because they are not important for our application. From Figure \ref{fig:zernike_results}, we can see that the results of our method are in agreement with Zemax OpticStudio. All residuals are less than $\lambda/23$.

Our method offers several advantages over the methods discussed in Section \ref{sec:background}. Our method works with off-axis field angles, fast beams, irregular obscurations, and arbitrary reference wavefronts. In addition, unlike the classical Hartmann test approaches discussed in Section \ref{sec:classical_Hartmann}, we do not need to know the actual distribution of the holes on the Hartmann mask in our method.

Our method has been demonstrated on simulated data from the G-CLEF red camera test setup. At the time of writing, we have started to the implement the test setup in lab, as seen in Figures \ref{fig:Red_AIT_lab_layout} and \ref{fig:Red_AIT_lab_Hartmann}. The next step would be to verify our method on real-world data taken using our test setup in lab.

%% file: Sections/conclusions.tex
\section{Conclusions}

We developed a wavefront sensing method which generalizes the classical Hartmann test for off-axis field angles, allowing us to quantify the aberrations in a focal optical system when light is injected off-axis. Our method works with fast beams, arbitrary reference wavefronts, and irregular obscurations. In our method, we take images at two defocused observation planes, so that we can trace real rays back to the reference wavefront (Sections \ref{sec:TRA_centroids} and \ref{sec:TRA_trace_back}). We then propagate ideal rays forward from the reference wavefront to one of the two observation planes, and compute transverse ray aberrations on that plane (Section \ref{sec:TRA_prop_forward}). We derived a pair of nonlinear PDEs relating transverse ray aberrations and wavefront error (Equation (\ref{eq:TRA_WFE_PDE})), and then solved these PDEs using Zernike decomposition and nonlinear least squares (Section \ref{sec:solve}), allowing us to estimate wavefront error. Our method has been demonstrated and verified on simulated data from G-CLEF's 7-lens $f$/2.25 red camera. We have started to implement a test setup in lab, and the next step is to verify our method experimentally.

%% file: Sections/appendix.tex
\appendix
\numberwithin{equation}{section}

\section{Intersection of a Line and a Sphere}\label{sec:app_line_sphere}

Consider a line and a sphere in $\mathbb{R}^3$. Let $(x_1,y_1,z_1)$ and $(x_2,y_2,z_2)$ be points on the line. The parameteric equations of the line are:
\begin{align}\label{eq:app_line}
    x &= x_1 + (x_2 - x_1)t \nonumber \\
    y &= y_1 + (y_2 - y_1)t \nonumber\\
    z &= z_1 + (z_2 - z_1)t
\end{align}
Let $(c_x,c_y,c_z)$ be the center of the sphere, and let $R$ be the radius of the sphere. The equation of the sphere is:
\begin{equation}\label{eq:app_sphere}
    (x-c_x)^2 + (y-c_y)^2 + (z-c_z)^2 - R^2 = 0
\end{equation}
By substituting Equation (\ref{eq:app_line}) into Equation (\ref{eq:app_sphere}) and collecting terms of $t$, the following quadratic equation can be obtained:
\begin{equation}\label{eq:app_subbed}
    at^2 + bt + c = 0
\end{equation}
where $a$, $b$, and $c$ are constants, defined as:
\begin{align}
    a &\equiv (x_1-x_2)^2 + (y_1-y_2)^2 + (z_1-z_2)^2 \nonumber\\
    b &\equiv 2 \left[(c_x-x_1)(x_1-x_2) + (c_y-y_1)(y_1-y_2) + (c_z-z_1)(z_1-z_2)\right] \nonumber\\
    c &\equiv (c_x-x_1)^2 + (c_y-y_1)^2 + (c_z-z_1)^2 - R^2
\end{align}
We seek a value of $t$, called $t^\star$, such that Equation (\ref{eq:app_subbed}) is satisfied. This can be found with the quadratic formula:
\begin{equation}
    t^\star = \frac{-b \pm \sqrt{b^2 - 4ac}}{2a}
\end{equation}
If a solution exists for $t^\star$, then the point of intersection between the line and the sphere can be found by substituting $t^\star$ into Equation (\ref{eq:app_line}). Note that there can be two points of intersection, one point of intersection, or no points of intersection, depending on the value of the discriminant $b^2 - 4ac$.


\section{Zernike Circle Polynomials}\label{sec:app_Zernike}

\begin{figure}[H]
  \centering
  \includegraphics[width=\textwidth]{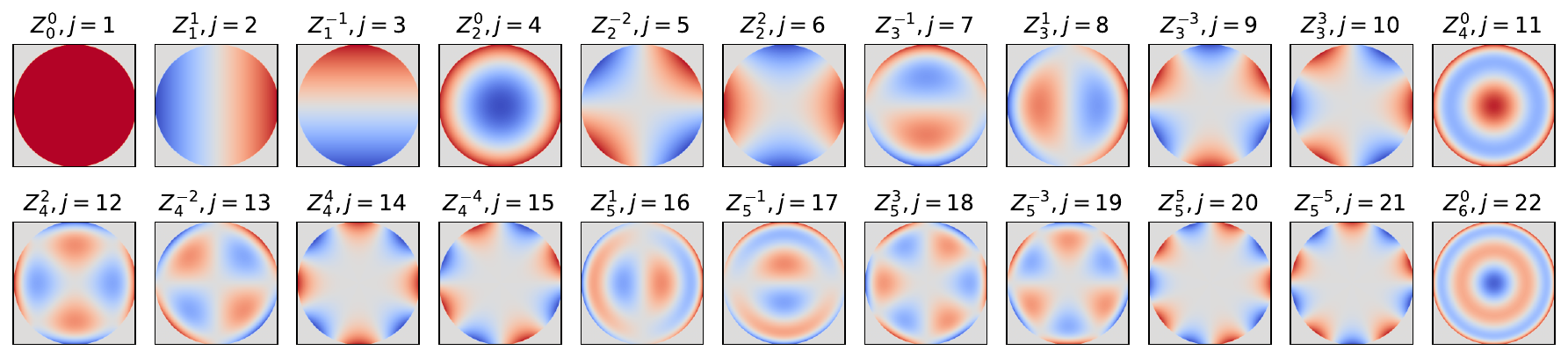}
  \caption{The first 22 Zernike circle polynomials according to the Noll indexing scheme. $j$ is the Noll Zernike index. Red values are positive and blue values are negative.}
  \label{fig:noll_zernike}
\end{figure}

The Zernike circle polynomials\cite{Zernike1934} are a set of orthogonal polynomials on the unit disk. Because of their orthogonality on the unit disk, they are commonly used to decompose the wavefront error of an optical system with a circular aperture. Each Zernike polynomial represents a particular kind of optical aberration (e.g. tilt, defocus, astigmatism, coma, trefoil, spherical aberration). Ref. \citenum{Sch2016} defines the Zernike circle polynomials in polar coordinates as:
\begin{equation}
    Z^m_n(r,\phi) \equiv 
    \left\{
    \begin{aligned}
        N^m_n R^{|m|}_n(r) \cos{(m\phi)} &\:  \text{ if } m \geq 0\\
        N^m_n R^{|m|}_n(r) \sin{(m\phi)} &\: \text{ if } m < 0
    \end{aligned}
    \right.
\end{equation}
where $n$ and $m$ index each polynomial. $n$ is a nonnegative integer and $m$ can take on values $-n, -n+2, -n+4, \ldots, n$. $R^{|m|}_n(r)$ are the radial polynomials, defined as:
\begin{equation}
    R^{|m|}_n(r) \equiv \sum^{(n-|m|)/2}_{k=0} (-1)^k \binom{n-k}{k} \binom{n-2k}{(n-m)/2-k}
\end{equation}
$N^m_n$ is a normalization constant, defined so that $\int_0^1 \int_0^{2\pi} Z^m_n(r,\phi) \,d\phi \,dr = 1$. $N^m_n$ is defined as:
\begin{equation}
    N^m_n \equiv \sqrt{\frac{2(n+1)}{1+\delta_{m,0}}}
\end{equation}
where $\delta$ is the Kronecker delta. The partial derivatives of the Zernike polynomials in Cartesian coordinates can also be expressed as a linear combination of Zernike polynomials\cite{Sch2016,Stephenson2014}. Ref. \citenum{Sch2016} gives this as:
\begin{multline}\label{eq:dZ_dx}
    \frac{\partial Z^m_n}{\partial x} = \sqrt{(1+\delta_{m,0})(n+1)}
    \Biggl( \sum^{n-1}_{\substack{n'=|m|+1 \\ \mathrm{Step\:2}}} \sqrt{n'+1} \: Z_{n'}^{\frac{m}{|m|}(|m|+1)}
    + (1-\delta_{m,0})(1-\delta_{m,-1})
    \\ \times \sqrt{1+\delta_{m,1}} 
    \sum^{n-1}_{\substack{n'=|m|-1 \\ \mathrm{Step\:2}}} \sqrt{n'+1} \: Z_{n'}^{\frac{m}{|m|}(|m|-1)}
    \Biggr)
\end{multline}
\begin{multline}\label{eq:dZ_dy}
    \frac{\partial Z^m_n}{\partial y} = \sqrt{(1+\delta_{m,0})(n+1)} \:\frac{m}{|m|}
    \Biggl( \sum^{n-1}_{\substack{n'=|m|+1 \\ \mathrm{Step\:2}}} \sqrt{n'+1} \: Z_{n'}^{-\frac{m}{|m|}(|m|+1)}
    - (1-\delta_{m,0})(1-\delta_{m,1})
    \\ \times \sqrt{1+\delta_{m,-1}} 
    \sum^{n-1}_{\substack{n'=|m|-1 \\ \mathrm{Step\:2}}} \sqrt{n'+1} \: Z_{n'}^{-\frac{m}{|m|}(|m|-1)}
    \Biggr)
\end{multline}
Note that the summations have a step size of 2. In practice, conventionally, each Zernike polynomial is referred to by a single index. There are several indexing schemes which map a pair of allowed $n$ and $m$ values to a single unique nonnegative number $j$. In this work, the Noll indexing scheme \cite{Noll1976} was used. The first 22 Zernike polynomials according to the Noll indexing scheme are shown in Figure \ref{fig:noll_zernike}.

